\documentclass[aps,prx,twocolumn,showpacs,amsmath,amssymb,floatfix,superscriptaddress,notitlepage]{revtex4-2}

\usepackage{lipsum,color,bbm,graphicx}
\usepackage{dcolumn}
\usepackage[utf8]{inputenc}
\usepackage{epstopdf}
\usepackage{amsthm,amsmath,empheq,bbm,braket,amssymb}
\usepackage[usenames,dvipsnames]{xcolor}
\usepackage{cases}
\usepackage{latexsym}
\usepackage{enumitem}
\usepackage[colorlinks=true,citecolor=Cerulean,linkcolor=RubineRed,urlcolor=Cerulean]{hyperref}

\usepackage{float}

\usepackage{stackengine}

\stackMath
\usepackage[caption=false]{subfig}
\graphicspath{ {images/} }

\DeclareFontFamily{OT1}{pzc}{}
\DeclareFontShape{OT1}{pzc}{m}{it}{<-> s * [1.15] pzcmi7t}{}
\DeclareMathAlphabet{\mathpzc}{OT1}{pzc}{m}{it}

\newcommand{\id}{\mathbbm{1}} 
\newcommand{\tr}[1]{\operatorname{\textnormal{Tr}}\left( {#1} \right)} 

 \newcommand{\dt}{\hspace{1pt}dt\hspace{1pt}}
 
\newcommand{\cov}{\textnormal{cov}}

\newcommand{\rvec}{\boldsymbol{r}}

\newcommand{\Hmeas}{H_\textnormal{meas}}
\newcommand{\chron}{\mathcal{X}}%
\newcommand{\Hfb}{H_\textnormal{fback}^\chron}

\newcommand{\Prob}{\mathrm{P}}

\begin{document}

\title{
Reshaping the Quantum Arrow of Time
}

\author{Luis Pedro Garc\'ia-Pintos}
\email{lpgp@lanl.gov}
\affiliation{Quantum and Condensed Matter Physics Group (T-4), Theoretical Division, Los Alamos National Laboratory, Los Alamos, New Mexico 87545, USA}

\author{Yi-Kai Liu}
\affiliation{Joint Center for Quantum Information and Computer Science, NIST/University of Maryland, College Park, Maryland 20742, USA} 
\affiliation{Applied and Computational Mathematics Division, National Institute of Standards and Technology, Gaithersburg, Maryland 20899, USA}

\author{Alexey V. Gorshkov}
\affiliation{Joint Center for Quantum Information and Computer Science, NIST/University of Maryland, College Park, Maryland 20742, USA}
\affiliation{Joint Quantum Institute, NIST/University of Maryland, College Park, Maryland 20742, USA} 

\date{\today}

\begin{abstract}
While the microscopic laws of physics are often symmetric under time reversal, most natural processes that we observe are not. The emergent asymmetry between typical and time-reversed processes is referred to as the arrow of time. In quantum physics, an arrow of time emerges when a sequence of measurements is performed on a system. We introduce quantum control tools that can yield dynamics more consistent with time flowing backward than forward. The control tools are based on the explicit construction of a Hamiltonian that can replicate the stochastic trajectories of a monitored quantum system. Such Hamiltonian can reverse the effect of monitoring and, via a feedback process, generate trajectories consistent with a reversed arrow of time. It can also be used to simulate the backward-in-time dynamics of an open quantum system. Finally, we design a feedback-driven continuous measurement engine powered by the energy pumped into the system by the monitoring process. We show the engine can operate under experimentally realizable conditions with feedback delay and finite-efficiency measurements. 
\end{abstract}

\maketitle

Scientists and philosophers refer to the time asymmetry observed in natural phenomena as the \emph{arrow of time}. Most natural processes we observe follow an arrow of time: we see eggs hatching, glasses breaking, and stars exploding, but never their time-reversed processes. This is so even though most microscopic laws of physics are symmetric under time reversal (with the notable exception of the weak nuclear force~\cite{lees2012observation}).

We call \emph{forward} processes those that causally occur in nature, and \emph{backward} processes their time-reversed versions. In the latter, a system starts from the final state of a forward process and evolves under the same Hamiltonian but with quantities odd under time reversal negated~\cite{Timereversal} (e.g., momenta and magnetic fields change sign, but not positions). Backward processes are consistent with the laws of physics~\cite{arrowlamb,Timereversal}. How a direction to the arrow of time emerges from time-symmetric laws has puzzled scientists and philosophers for decades~\cite{prigogine2000arrow}.

Arrows of time have been attributed to many different origins~\cite{BookZeh1989direction, BookPrice1996time, BookHalliwell1996physical, BookCarroll2010eternity}. In cosmology, for instance, a particular low-entropy initial state of the universe can lead to a cosmological arrow of time~\cite{ArrowIntitial1975, CosmoArrowHawking1985, ellis2013arrow}. Other possible characterizations of emergent arrows of time have been attributed to gravity~\cite{gravityarrow}, quantum entanglement~\cite{Zurek_1998, RudolphPRE2010}, one's perceptions~\cite{10.1162/netn_a_00300, wolpert2024memory}, computer science arguments~\cite{arrighi}, and even to modifications to known physical laws~\cite{Penrose:1980ge,prigogine2000arrow}.

Perhaps the most recognizable arrow of time is the thermodynamic one, which arises despite the time symmetry of the underlying microscopic dynamics behind thermodynamic processes. The second law of thermodynamics dictates that, for macroscopic systems, processes that increase entropy are overwhelmingly more likely than those that decrease entropy, and on average, the universe's entropy increases. The larger the system, the harder it is to observe anomalous, entropy-decreasing dynamics. The manifestation of the thermodynamic arrow of time can be quantified by comparing the likelihood of a process with its time reverse~\cite{CrooksArrow, Jarzynski,
seif2021machine}. Stochastic fluctuations inherent to thermodynamic processes can signal the emergence of a thermodynamic arrow of time.

While classical randomness is understood to derive from a lack of complete knowledge of the microscopic description of a system, quantum randomness in measurement outcomes is, to the best of our knowledge, fundamental~\cite{pironio2010random, acin2016certified}. The most complete description of a quantum system, its wavefunction, only yields probabilities of possible measurement outcomes. The fundamentally stochastic dynamics of measured quantum systems leads to an arrow of time that can be quantified similarly to the thermodynamic one, as shown by Ref.~\cite{ArrowDresselPRL2017}. 

Here, we introduce quantum control tools that can manipulate such a quantum arrow of time. We provide an explicit construction of a Hamiltonian that can replicate the stochastic dynamics of a measured quantum system. We will see that, by suitable feedback, such a Hamiltonian can generate trajectories consistent with a modified arrow of time. An agent that can perform feedback can stretch the perceived arrow of time or even invert its direction, challenging the notion that the arrow of time can always be determined unambiguously by an observer. The agent can generate other anomalous physical processes, such as driving a measurement engine powered by the monitoring process.

\section{Background: The length of the quantum arrow of time}
\label{sec:Background}

Upon observing a measurement outcome $r$ that occurs with probability $P(r|\rho)$, the state $\rho$ of a quantum system transforms by 
 \begin{align}
 \label{eq-measurement}
 \rho \,\, \xrightarrow[]{\,\text{outcome} \,\, r \,} \,\, \rho' = \frac{M_r \rho M_r^\dag}{P(r|\rho)}.
 \end{align}
 The $M_r$s are Kraus operators that characterize the post-measurement states, with $\int M_r^\dag M_r dr = \id$. Equation~\eqref{eq-measurement} can describe strong projective measurements (when the $M_r$s are projectors) and generalized measurements in which one only gains partial information of the system's state (e.g., by strongly measuring an apparatus that is correlated with the system of interest)~\cite{Bookwiseman2009}. We assume pure states and use the system's wavefunction $\ket{\psi}$ or density matrix $\rho \coloneqq \ket{\psi}\!\bra{\psi}$ interchangeably to denote the system's state. (We will only assume mixed states in Sec.~\ref{sec:Applications}, where we consider ensemble averages and study applications with non-ideal feedback processes.) 
 
 Throughout this work, we assume generalized Gaussian measurements on systems of arbitrary dimension occurring over short times $dt$, with a distribution of outcomes
 \begin{align}
 \label{eq:probsingle}
 P(r|\rho) \nobreak  = \tr{\rho M_r^\dag M_r} = \nobreak \sqrt{\frac{dt}{2\pi\tau}} e^{ -\left(r-\langle A \rangle \right)^2 dt/(2\tau) }.
 \end{align}
 $A$ and $\langle A \rangle \coloneqq \tr{A \rho}$ denote the measured observable and its expectation value evaluated at the system's state $\rho$ when the measurement occurs. The parameter $\tau$, referred to as the characteristic measurement time, determines the outcomes' standard deviation $\sqrt{\tau/dt}$. In a procedure where a measurement apparatus interacts with the system during a time $dt$, $\tau$ depends on the strength of the apparatus-system coupling and dictates how long it takes for a sequence of weak measurements to approximate a strong projective measurement~\cite{JacobsIntro2006}.
 
 The measurement output $r$ tracks the monitored observable's expectation value $\langle A \rangle$, with $r dt \nobreak =\nobreak \langle A \rangle dt \nobreak+\nobreak  \sqrt{\tau}dW$, where $dW$ denotes stochastic Wiener noise~\citep{JacobsIntro2006}. Then, upon averaging the results of $r$ over many realizations of a generalized measurement on the same state, one recovers $\overline{r} = \langle A \rangle$. $\overline{f}$ denotes the average over many measurement instances of a quantity $f$, not to be confused with the quantum mechanical expectation value $\langle A \rangle$ of the observable.
 
 The Kraus operators $M_r\nobreak \coloneqq \nobreak \left(dt/2\pi\tau \right)^{1/4} e^{ -(r-A)^2 dt/(4\tau) }$ describe such Gaussian measurements~\cite{Bookjacobs2014}. These provide a good description of, e.g.,  measurements on superconducting qubits~\cite{superconducting1, superconducting2, PhysRevX.6.011002}.
 If one focuses on observables that satisfy $A^2 = \id$ (as we describe below, this assumption implies the backward process is a possible physical process), e.g., for strings of Pauli matrices in the case where our system consists of qubits, the Kraus operators simplify to 
 \begin{align}
 \label{eq:Kraus}
 M_r\nobreak  \propto \nobreak  e^{ + r A dt/(2\tau) }.
 \end{align}
 The other terms in the exponential of $M_r$ do not affect the system's state but are instead absorbed by the Kraus operators' normalization.

Consider a system evolving under a Hamiltonian $H$ that undergoes a sequence of measurements as in Eq.~\eqref{eq-measurement}. The system goes through states $\ket{\psi}_{t_j}$ at times $t_j \nobreak = \nobreak j dt$, $j=\{0,\ldots,N\}$, which are determined by $H$ and the previous measurement outcomes. We denote the outcomes by $\rvec \nobreak=\nobreak \{r_{t_0}, \ldots r_{t_j}, \ldots r_T \}$. Assuming that $dt \ll \{1/\|H\|,\tau\}$ is smaller than all other timescales, the final state at time $T =  N  dt$ in the \emph{forward process} is 
  \begin{align} \label{eq-forward}
 \ket{\psi}_{T} = \frac{ M_{r_T} e^{-i H dt} \ldots 
\overbrace{M_{r_{t_j}}}^{\color{blue}\text{\clap{output correlated with $\langle A \rangle$}}}
 e^{-i H dt} \ldots M_{r_{t_0}} e^{-i H dt} \ket{\psi}_{0} }{\sqrt{\Prob_F(\rvec)}}.
 \end{align}
$\Prob_F(\rvec) \equiv \prod_{j=1}^N P\big(r_{t_j}\big\vert\ket{\psi}_{t_j}\!\big)$ is the probability with which the sequence of outcomes $\rvec$ occurs, determined by the state's trajectory. (Since $dt$ is small, ordering the operators as $M_{r_j} e^{-iHdt}$ or $e^{-iHdt} M_{r_j}$ 
in Eq.~\eqref{eq-forward} yields the same trajectory, to leading order~\cite{JacobsIntro2006}.)

The \emph{backward process} exactly retraces the trajectory in Eq.~\eqref{eq-forward} backward in time. We define it as the time-reversed (and outcome-negated) version of Eq.~\eqref{eq-forward}:
 \begin{align}
 \label{eq-backward}
 \ket{\psi}_{0} = \frac{ e^{-i \widetilde{H} dt} M_{-r_{t_0}}  \ldots  e^{-i \widetilde{H} dt} 
 \overbrace{M_{-r_{t_j}}}^{\color{red}\text{\clap{output \emph{anticorrelated} with $\langle A \rangle$}}}
  \ldots e^{-i \widetilde{H} dt}  M_{-r_T}  \ket{\psi}_{T}   }{\sqrt{\Prob_B(\rvec)}}.
 \end{align}
 As shown in Ref.~\cite{ArrowDresselPRL2017}, the backward process is also consistent with the laws of physics under the appropriate transformations. The operations that lead to a physical time-reversed trajectory involve negating measurement outcomes $\rvec \rightarrow -\rvec$~\cite{ArrowDresselPRL2017} and an operation $H \rightarrow \widetilde{H}$ that negates odd Hamiltonian variables~\cite{Timereversal} (e.g., flipping magnetic fields and momenta but not positions). To see that Eq.~\eqref{eq-backward} is consistent with physical laws, it suffices to note that, if $\rho' = M_r \rho M_r^\dag/P_f$ for a forward measurement process, it also holds that $\rho = M_{-r} \rho' M_{-r}^\dag/P_b$ from Eq.~\eqref{eq:Kraus} (albeit with different probabilities $P_f$ and $P_b$). That is, a time-reversed trajectory that takes the final post-measurement state $\rho'$ to the initial pre-measurement one $\rho$, as in Eq.~\eqref{eq-backward}, is physically possible.

However, the probabilities $\Prob_F(\rvec)$ and $\Prob_B(\rvec)$ with which the forward trajectory in Eq.~\eqref{eq-forward} and its time-reversed version~\eqref{eq-backward} occur are different. This is because, while $\rvec$ in the forward process tracks the observable's expectation value $\langle A \rangle$, $-\rvec$ in the backward process is anti-correlated with the observable. The relative likelihood $\mathcal{R}$ of the forward and backward processes satisfies
\begin{align}
\label{eq:arrow}
\ln   \mathcal{R} \, \stackrel{\text{(i)}}{\coloneqq} \, \ln \frac{\Prob_F(\rvec)}{\Prob_B(\rvec)} \, \stackrel{\text{(ii)}}{=} \, \frac{2}{\tau} \int_0^T r_t \langle A \rangle_t  \dt.
\end{align}
(i) in Eq.~\eqref{eq:arrow} was considered by Ref.~\cite{ArrowDresselPRL2017} as a quantifier of the arrow's length and direction, mirroring the work of Ref.~\cite{CrooksArrow} on the thermodynamic arrow of time. (ii) was proven for a qubit in Ref.~\cite{ArrowDresselPRL2017} (for completeness, we include the proof for any $A$ with $A^2 = \id$ in Appendix~\ref{app-quantumarrow}~\footnote{(ii) in Eq.~\eqref{eq:arrow} corresponds to a Stratonovich integral in stochastic calculus~\cite{jacobs2010stochastic,gardinerstochastic}: 
$\int_0^T r_t f_t dt \equiv \lim_{dt \rightarrow 0} \sum_j^N r_{t_{j}} dt \big( f_{t_j} + f_{t_{(j+1)}} \big)/2$.}.)
We use $r_t$, $\rho_t$, and $\langle A \rangle_t$ to denote the time-continuous versions of the measurement outcomes, state, and observable's expectation value.

We stress that Eq.~\eqref{eq:arrow} is not to be interpreted as an explanation for the origin of time's arrow. Instead, $\ln \mathcal{R}$ merely \emph{quantifies} time asymmetry~\cite{CrooksArrow, Jarzynski}. When $\ln \mathcal{R} \nobreak=\nobreak 0$, the forward trajectory and its time reverse are equally likely to have occurred.  Trajectories for which $\ln \mathcal{R} > 0$ are more likely to occur naturally by forward processes, i.e., they are more consistent with the standard direction for time's arrow. As these are probabilistic processes, $\ln \mathcal{R} < 0$ can occur but, typically, $\ln \mathcal{R} > 0$.

If a cheeky experimentalist hands us the measurement output and a trajectory ($\rvec$ and the history of $\langle A \rangle_t$) and their backward version, Eq.~\eqref{eq:arrow} can be used to (probabilistically) decide which of the two is the reversed one. Since $r_t$ is positively correlated with $\langle A \rangle_t$, the longer the run, the easier it is to decide which is the forward, causally generated trajectory~\cite{ArrowDresselPRL2017,ArrowMurchPRL2019, ArrowBigelowNatComm2021}. Similar figures of merit have been used to probe dissipation and time-reversal symmetry in, e.g., biological systems~\cite{collin2005verification,roldan2021quantifying}. 

An arrow of time emerges despite the time-reversal symmetry of the monitored dynamics. Its length and direction are quantified by Eq.~\eqref{eq:arrow}.
 Next, we introduce experimentally realizable quantum control tools that can blur time's arrow.

\section{Replicating stochastic quantum trajectories}

In the time-continuous limit, Eq.~\eqref{eq-forward} is equivalent to the difference equation 
\begin{align}
\label{eq:Strato}
d\rho_t &= -i[H,\rho_t]dt + \frac{r_t}{2\tau}  \Big( \big\{\rho_t,A\big\} - 2 \langle A \rangle_t \rho_t \Big) dt \\
&\qquad \qquad \quad \quad  - \frac{1}{4\tau} \Big( \big\{\rho_t,A^2\big\} - 2 \langle A^2 \rangle_t \rho_t \Big) dt  \nonumber 
\end{align} 
for the change in $\rho_t \coloneqq \ket{\psi}_t\!\bra{\psi}$.
We prove this in Appendix~\ref{app-stratonovich}. Equation~\eqref{eq:Strato} is to be interpreted as a Stratonovich equation in stochastic calculus~\cite{GisinStrato,jacobs2010stochastic,ArrowDresselPRL2017} (a different expression for $d\rho_t$ holds in the It\^o picture~\cite{JacobsIntro2006}).

The first result of this work is that the Hamiltonian
\begin{align}
\label{eq:Hmeas}
\Hmeas \coloneqq -i \frac{r_t}{2\tau}  \left[\rho_t,A \right] +i \frac{1}{4\tau} \left[\rho_t,A^2 \right]
\end{align} 
yields the same dynamics as the stochastic contribution to the monitored dynamics, so that Eq.~\eqref{eq:Strato} can be written as $d\rho_t = -i \left[ H + \Hmeas , \rho_t \right] dt$.
That is, driving a system with $\Hmeas + H$ reproduces the dynamics of a monitored quantum system with a self-Hamiltonian $H$. We prove this claim in Appendix~\ref{app-replicating}. The Hamiltonian that reproduces a stochastic path is not unique [for instance, adding functions of $\rho$ to Eq.~\eqref{eq:Hmeas} retraces the same stochastic path]. Note that $\Hmeas$ in Eq.~\eqref{eq:Hmeas} generates nonlinear dynamics on $\rho_t$ of a double bracket flow form~\cite{BROCKETT199179, frank2022physics, hastings2022lieb, Gluza2024doublebracket}.

A reader may wonder how it is possible that a Hamiltonian can reproduce the fundamentally stochastic dynamics of a monitored quantum system. The ``catch'' is that $\Hmeas$ uses detailed information about a trajectory to reproduce it: knowledge of the measurement outcomes $r_t$ and the system's state $\rho_t$ is required.
Another reader may instead wonder whether the information needed to implement $\Hmeas$ can ever be available. The answer is: yes! Equations~\eqref{eq:Kraus} and~\eqref{eq-forward} provide an update rule that yields an exact trajectory given knowledge of a system's initial state, the Hamiltonian, and the measurement outcomes $\rvec$. 
For systems of sufficiently small dimension, the update rule allows one to reconstruct $\rho_t$ from the measurement record via classical simulation and, from it, to obtain $\Hmeas$. 

\begin{figure}  \centering    
(a) Original or copycat?
\includegraphics[trim=00 00 00 00,width=0.5 \textwidth]{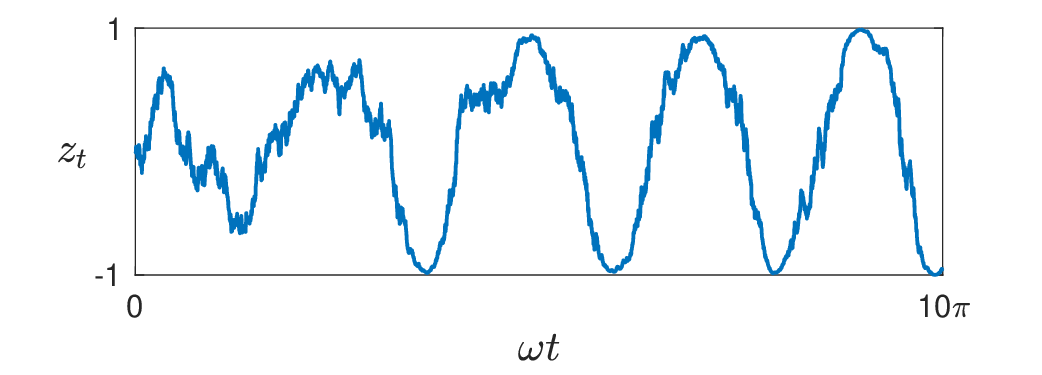} \\ 
(b) Original or copycat?
\includegraphics[trim=00 00 00 00,width=0.5 \textwidth]{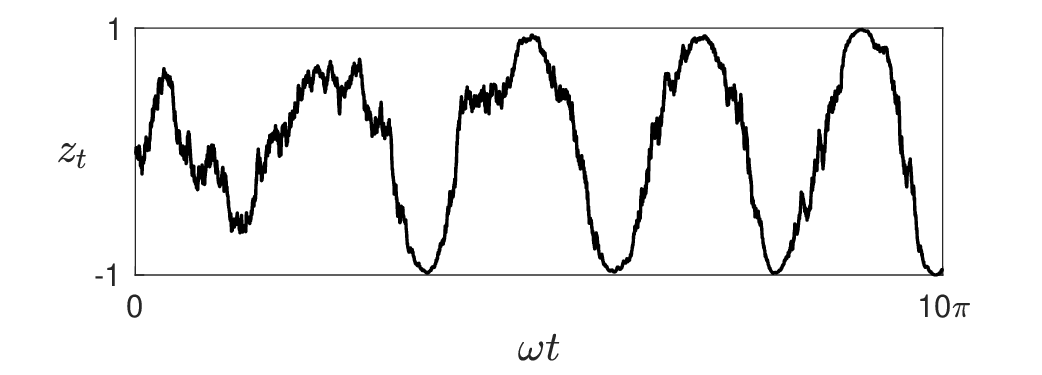} 
\caption{\label{fig:fig1} 
\textbf{Replicating stochastic quantum trajectories.} 
Simulations of the evolution of a monitored qubit's $z_t = \tr{\rho_t \sigma_z}$ component as a function of time. 
 For the simulations, $A = \sigma_z$, $H = \omega \sigma_y/2$, $\omega \tau = 2\pi$, and $\tau/dt = 10^3$. 
One of the two plots corresponds to the simulation of the stochastic dynamics of a monitored qubit. 
The other curve was generated by the Hamiltonian $\Hmeas$ in Eq.~\eqref{eq:Hmeas}. The stochastically generated original trajectory is reproduced by $\Hmeas$. 
[Plot (a) is the original stochastic trajectory and (b) is the one reproduced by $\Hmeas$.]
}
\end{figure}

That \emph{some} Hamiltonian can reproduce the trajectories is not surprising since the system goes through pure states. 
Reference~\cite{Minev_2019} exploits this fact to experimentally reverse stochastic jumps. 
References~\cite{alipourQuantum2020,alipourPRA2022,PhysRevResearch.5.033045} construct a Hamiltonian that reproduces arbitrary trace-preserving dynamics from information of the state dynamics. 
In the context of monitored systems, Ref.~\cite{hu2023describingwavefunctioncollapse} relies on the fact that unitaries connect states in a trajectory to show that the Hamiltonian $H_{\textnormal{HJ}} \nobreak\coloneqq\nobreak i \left( \ket{\tfrac{d}{dt}\tilde\psi_t}\bra{\tilde \psi_t} - \ket{\tilde \psi_t} \bra{\tfrac{d}{dt} \tilde \psi_t}\right) \nobreak + \nobreak \tfrac{d}{dt}\Phi_t \id$, where $\ket{\tilde\psi_t} = e^{i \Phi_t} \ket{\psi_t}$ and $\Phi_t = -i \int_0^t \bra{\tfrac{d}{dt'} \psi_{t'}} \psi_{t'} \rangle dt'$,
retraces the stochastic path followed by the system's state $\ket{\psi_t}$. Reference~\cite{GarrahanPRB2025} maps the stochastic trajectories in monitored circuits to explicit unitaries. 
The novelty of $\Hmeas$ in Eq.~\eqref{eq:Hmeas} lies in the explicit recipe it provides for reproducing the stochastic trajectory followed by a monitored quantum system given knowledge of the initial state and measurement output.

For example, consider the continuous monitoring of $A = \sigma_z$ on a qubit with a Hamiltonian $H = \omega \sigma_y/2$. If $\sigma_{\alpha = \{x,y,z\}}$ are the Pauli matrices, the qubit's state is $\rho_t = \left(\id + x_t \sigma_x + y_t \sigma_y  + z_t \sigma_z \right)/2$, where $x_t = \tr{\rho_t \sigma_x}$, $y_t = \tr{\rho_t \sigma_y}$, and $z_t = \tr{\rho_t \sigma_z}$ are the Bloch sphere coordinates. The Hamiltonian in Eq.~\eqref{eq:Hmeas} becomes $\Hmeas \nobreak = \nobreak - \frac{r_t}{2\tau} \left( x_t\sigma_y - y_t \sigma_x \right)$ (the same expression is obtained from the $H_{\textnormal{HJ}}$ in the previous paragraph~\cite{hu2023describingwavefunctioncollapse}). Knowledge of $\rho_0$ and $r_t$ allows calculating $x_t$ and $y_t$, yielding an easy-to-implement $\Hmeas$ that reproduces the qubit's monitored dynamics. 
We illustrate this in Fig.~\ref{fig:fig1}. 

Next, we show how to leverage $\Hmeas$ to influence the quantum arrow of time.

\section{Reshaping the quantum arrow of time}
\label{sec:Reshaping}

\begin{figure*}[ht!]\subfloat{\begin{tabular}[b]{c} {(a) Standard arrow of time ($\chron = 0$)}  \\  \includegraphics[width=0.9\columnwidth]{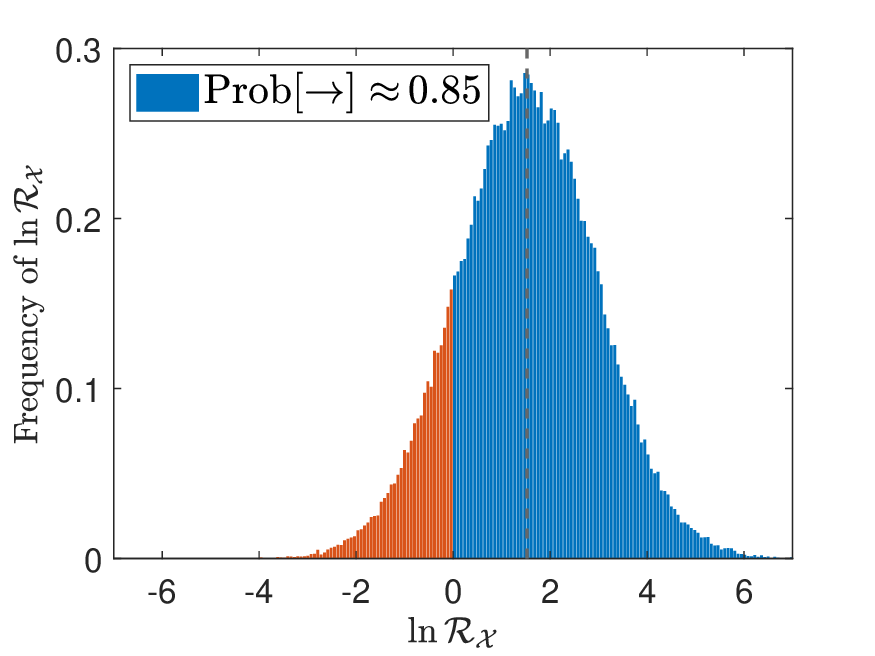}   \end{tabular}} \subfloat{\begin{tabular}[b]{c} {(b) Stretched arrow of time ($\chron = 1$)}  \\ \includegraphics[width=0.9\columnwidth]{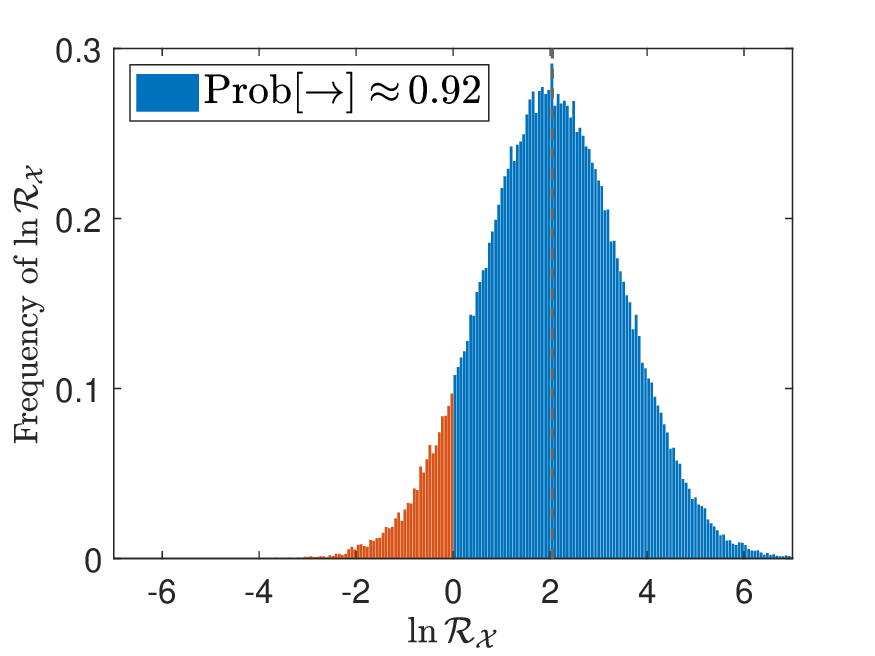}   \end{tabular}} 
 \\ \subfloat{\begin{tabular}[b]{c} {(c) Blurred arrow of time ($\chron = -3$)}  \\  \includegraphics[width=0.9\columnwidth]{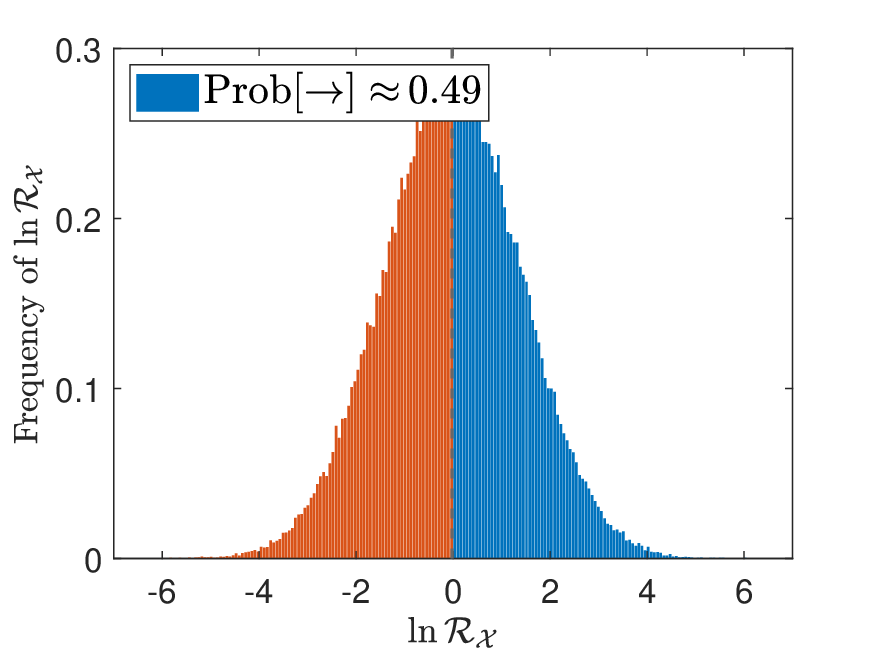}   \end{tabular}} 
\subfloat{\begin{tabular}[b]{c} {(d) Inverted arrow of time ($\chron = -4$)}  \\  \includegraphics[width=0.9\columnwidth]{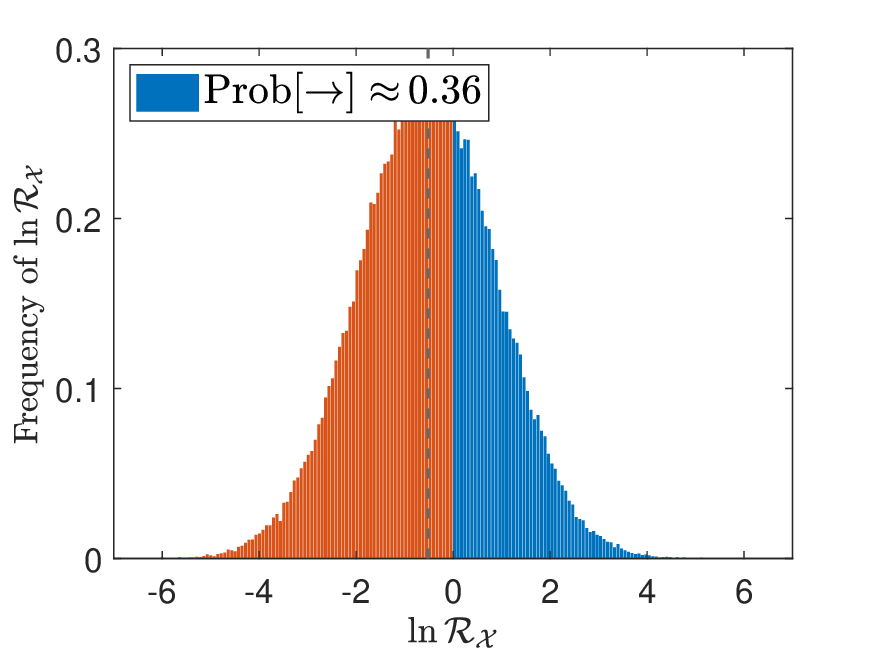}   \end{tabular}} \caption{ \label{fig:fig2}
\textbf{Reshaping the quantum arrow of time.} Normalized histograms of the log relative likelihood $\ln \mathcal{R}_\chron$ for $10^6$ realizations of stochastic trajectories for different values of the arrow-modifying factor $\chron$.  We consider a qubit where $A = \sigma_z$ is continuously monitored for a time
$T = \tau$, with $\omega \tau = 8\pi$ and $\tau/dt = 10^3$.  The areas of the blue regions characterize the fraction of trajectories $\textnormal{Prob}[\rightarrow]$ for which a forward arrow of time is more consistent than a backward one.  For $\chron = \{-4,-3,0,1\}$, the simulations show the respective values $\overline{ \ln \mathcal{R}_\chron} \approx \{-0.516, -0.022, 1.524, 2.046\}$, illustrated with vertical dashed lines in the histograms. Equation~\eqref{eq:arrow-steeredSimple} yields similar values of $\overline{ \ln \mathcal{R}_\chron} \approx \{-0.5,0,1.5,2\}$. While $\chron = 1$ stretches time's arrow [plot (b)], $\chron = -3$ prevents the arrow from manifesting [plot(c)], and $\chron = -4$ reverses its direction [plot (d)]. 
} 
\end{figure*}

Since $\Hmeas$ reproduces the monitored dynamics of a quantum system, sufficiently fast feedback with $-\Hmeas$ counteracts the measurement back-action. 
That is, replacing $H$ by $H - \Hmeas$ in Eq.~\eqref{eq:Strato} recovers the dynamics the system would have had if no monitoring had taken place, where $d\rho_t = -i[H,\rho_t]dt$. 
It is suggestive to explore the influence of such feedback processes on time's arrow. 

We will consider two different uses of $\Hmeas$ to influence the perceived flow of time: 
 (a, this section) feedback based on $\Hmeas$ to generate trajectories with a modified arrow of time, and (b, next section) evolving a system with $-H -\Hmeas$ to emulate backward-in-time dynamics. (From now on, we consider $A^2 = \id$, in which case backward processes are physically possible, as described in Sec.~\ref{sec:Background}.)  
 
Consider the more general feedback Hamiltonian, 
\begin{align}
\label{eq:Hfb}
\Hfb \coloneqq \chron \Hmeas = \chron \left( -i \frac{r_t}{2\tau}  \left[\rho_t,A \right] \right),
\end{align}
which acts immediately after a measurement instance with outcome $r_t$, where $\chron \in \mathbb{R}$. The measurement-plus-feedback effect's on the forward process involves transformations $ \ket{\psi}_{t_{j+1}} \nobreak \propto \nobreak e^{-i  \Hfb dt}  M_{r_{t_j}} \ket{\psi}_{t_{j}}$. From Eq.~\eqref{eq:Kraus} and the fact that $\Hmeas$ has the same effect as $M_r$, one concludes that, for positive $\chron$, the feedback effectively speeds up the monitoring process by decreasing the characteristic measurement time to $\tau/\chron$ [see the discussion after Eq.~\eqref{eq:probsingle}]. 

Meanwhile, the backward process involves transitions  
\begin{align}
\label{eq:BackwardFB}
\ket{\psi}_{t_{j}} &\propto M_{-r_{t_j}} e^{-i  \widetilde \Hfb dt}   \ket{\psi}_{t_{j+1}} \nonumber \\
&\propto  e^{-r_{t_j} A dt/(2\tau)} e^{+i \chron \Hmeas} \ket{\psi}_{t_{j+1}} \nonumber \\
&\propto   e^{-r_{t_j} A dt/(2\tau)} e^{-\chron r_{t_j} A dt/(2\tau)} \ket{\psi}_{t_{j+1}}. 
\end{align}
To derive Eq.~\eqref{eq:BackwardFB},
we used: 
\begin{itemize}
    \item[(i)] Equation~\eqref{eq:Kraus},
    \item[(ii)] that the measurement output in the Kraus operator is negated in the backward process,
    \item[(iii)] that fields in $\Hfb$ are negated in the backward process [equivalent to negating the measurement outcome from Eq.~\eqref{eq:Hfb}], and
    \item[(iv)] that $\Hmeas$ in Eq.~\eqref{eq:Hmeas} over an interval $dt$ induces the same dynamics as the effect of observing an outcome $r_{t_j}$, as we showed in the previous section.
\end{itemize}
For $\chron = -2$, Eq.~\eqref{eq:BackwardFB} implies that $\ket{\psi_{t_{j}}} \propto e^{r_{t_{j}}A\dt/(2\tau)} \ket{\psi_{t_{j+1}}} \propto M_{r_{t_j}}\ket{\psi_{t_{j+1}}}$, 
so that feedback 
makes the backward process resemble a forward process. This further suggests that $\chron$ influences time's arrow.

We prove in Appendix~\ref{app-steeringarrow} that, under such feedback process, the relative likelihood $\mathcal{R}_\chron \nobreak \coloneqq \nobreak  \Prob^\chron_F(\rvec)/\Prob^\chron_B(\rvec) $ between forward and backward processes satisfies 
\begin{align}
\label{eq:steeredarrow}
\ln   \mathcal{R}_\chron \, = \frac{2}{\tau} \int_0^T r_t \langle A \rangle_t dt = \, \ln \mathcal{R} +  \frac{\chron}{\tau}\int_0^T \! \Big(1-\langle A \rangle_t^2 \Big) \, dt.
\end{align}
$\Hfb$ modifies the effective arrow of time, generating trajectories with a varied degree of agreement with the standard, forward arrow of time. The free parameter $\chron$ governs the influence on time's arrow. Since $A^2 = \id$, it holds that $\int_0^T \left( 1 - \langle A \rangle^2 \right) dt \geq 0$. Thus, the sign of the second term in Eq.~\eqref{eq:steeredarrow} depends on the sign of $\chron$. Positive $\chron$ stretches time's arrow, while $\chron < 1$ squeezes it or, as we show next, even reverses its direction for sufficiently large $-\chron > 0$.

To illustrate how the arrow of time is reshaped by feedback, consider a qubit with $A = \sigma_z$ ($z_t \coloneqq \langle A \rangle_t$), as in the previous section's example. Ref.~\cite{ArrowDresselPRL2017} shows that the average of Eq.~\eqref{eq:arrow} is $\overline{\ln \mathcal{R}} = \int_0^T \overline{1 + z_t^2} dt/\tau$ and that $\int_0^T \overline{z_t^2} dt/T \approx 1/2$ when $T$ is a multiple of (or much larger than) $2\pi/\omega$ (we include a proof in Appendix~\ref{app-quantumarrow}). Using the two previous equalities, the ensemble average of Eq.~\eqref{eq:steeredarrow} is
\begin{align}
\label{eq:arrow-steeredSimple}
\overline{\ln \mathcal{R}_\chron} \approx \frac{T}{2\tau}\left( 3 + \chron  \right).
\end{align}
$\chron = 0$ recovers the standard length for time's arrow, which grows with the duration $T$ of the monitoring process. Feedback with $\chron > 0$ stretches time's arrow, generating trajectories where the forward and backward processes differentiate more than without feedback. In contrast, $-3 < \chron < 0$ shrinks time's arrow: the feedback makes it harder to discriminate forward and backward processes.

Perhaps the most counterintuitive regime occurs for $\chron \leq -3$, when $\overline{\ln \mathcal{R}_\chron}$ becomes zero ($\chron = -3$) or turns negative ($\chron < -3$). 
For $\chron = -3$, the feedback process generates trajectories without a clear direction for the arrow of time. Meanwhile, for $\chron <-3$, the trajectories are, on average, 
more consistent with a backward arrow of time. This happens because, as Eq.~\eqref{eq:BackwardFB} shows, values of $\chron < -2$ generate trajectories whose backward process resembles a forward process. For sufficiently negative $\chron$, the correlations between $\langle A \rangle$ and the backward process are higher than with the forward process. 
Contrary to what one may expect from the discussion after Eq.~\eqref{eq:Hfb}, a blurred arrow of time ($\overline{\ln \mathcal{R}_\chron} = 0$), does not occur at $\chron = -2$. This is because, even if the backward process resembles a forward one at $\chron = -2$, the measurement output is still causally generated from the forward one. Stronger feedback that causes  $\langle A \rangle$ to be more correlated with the backward process than the forward one is needed to reverse the direction of the arrow of time. 
We show examples of reshaped quantum arrows of time in Fig.~\ref{fig:fig2}.

That feedback can change the perceived arrow of time is not as surprising as one may initially think. Maxwell considered a being that, by a careful monitoring-and-feedback process, decreases the total entropy of two gases at different temperatures~\cite{Maxwell}. Maxwell's ``intelligent demon~\cite{LordKelvin}'' instigates a process that, while allowed by the laws of physics, is overwhelmingly unlikely to be observed in nature. The reverse of such process is the one we typically observe. That is, Maxwell's demon exploits measurement and feedback to generate a process consistent with a backward arrow of time. Similarly, the feedback Hamiltonian we introduced in Eq.~\eqref{eq:Hfb} serves as a control tool that can generate quantum trajectories more consistent with a backward quantum arrow of time.

The previous results open a rather provocative possibility. We have shown how the arrow of time can be manipulated, or even reversed, by an agent exploiting feedback control. Moreover, the feedback-driven trajectories are consistent with (feedback-less) backward trajectories, as described after Eq.~\eqref{eq:BackwardFB}. This allows for the possibility that nature could actually be tricking us about the direction of time. For instance, there is a way that nature could apply measurements and feedback control so that time seems to flow backward. Quantitative measures of the arrow of time, such as  Eq.~\eqref{eq:arrow}, would indicate that time flows backward. That is, in certain pathological cases, it is impossible to determine which direction the arrow of time is pointing. Even though there seems to be an emergent arrow of time, such an arrow is subjective, in the sense that its true direction is unknown and cannot be known (unless one could somehow detect the presence of external feedback mechanisms).

\section{Anomalous thermodynamic processes}
\label{sec:Applications}

We showed how to manipulate the perceived flow of time via suitably chosen feedback dynamics. One wonders how far these manipulations can go: can $\Hmeas$ and $\Hfb$ generate other anomalous physical processes? We explore this next. We show how to leverage $\Hmeas$ and $\Hfb$ to simulate the time-reversed dynamics of an open quantum system or continuously draw energy from the monitoring process.

\subsection{Simulating backward-in-time open dynamics}

The dynamics of a continuously monitored system averaged over runs of an experiment is equivalent to an open quantum system following Lindblad dynamics~\cite{Bookwiseman2009, Bookjacobs2014, JacobsIntro2006}. Let $\overline{\rho_{\xi,t}}$ denote averages over stochastic trajectories $\rho_{\xi,t}$ of pure states obtained from ideal continuous measurement processes. The index $\xi$ labels realizations of the stochastic trajectories. The ensemble-averaged mixed state $\overline{\rho_{\xi,t}}$ satisfies the Lindblad master equation $d\overline{\rho_{\xi,t}} \nobreak=\nobreak -i[H,\overline{\rho_{\xi,t}}]dt \nobreak-\nobreak \tfrac{dt}{8\tau} [A,[A,\overline{\rho_{\xi,t}}]]$~\cite{JacobsIntro2006}. Since $\Hmeas$ can reproduce individual stochastic trajectories [see discussion after Eq.~\eqref{eq:Hmeas}], averaging over realizations of $\Hmeas$ can emulate the dynamics of an open quantum system, as we show next.

\begin{figure}[hbt!]
 {\footnotesize (a) Simulating backward-in-time open dynamics }
\includegraphics[width=0.48\textwidth]{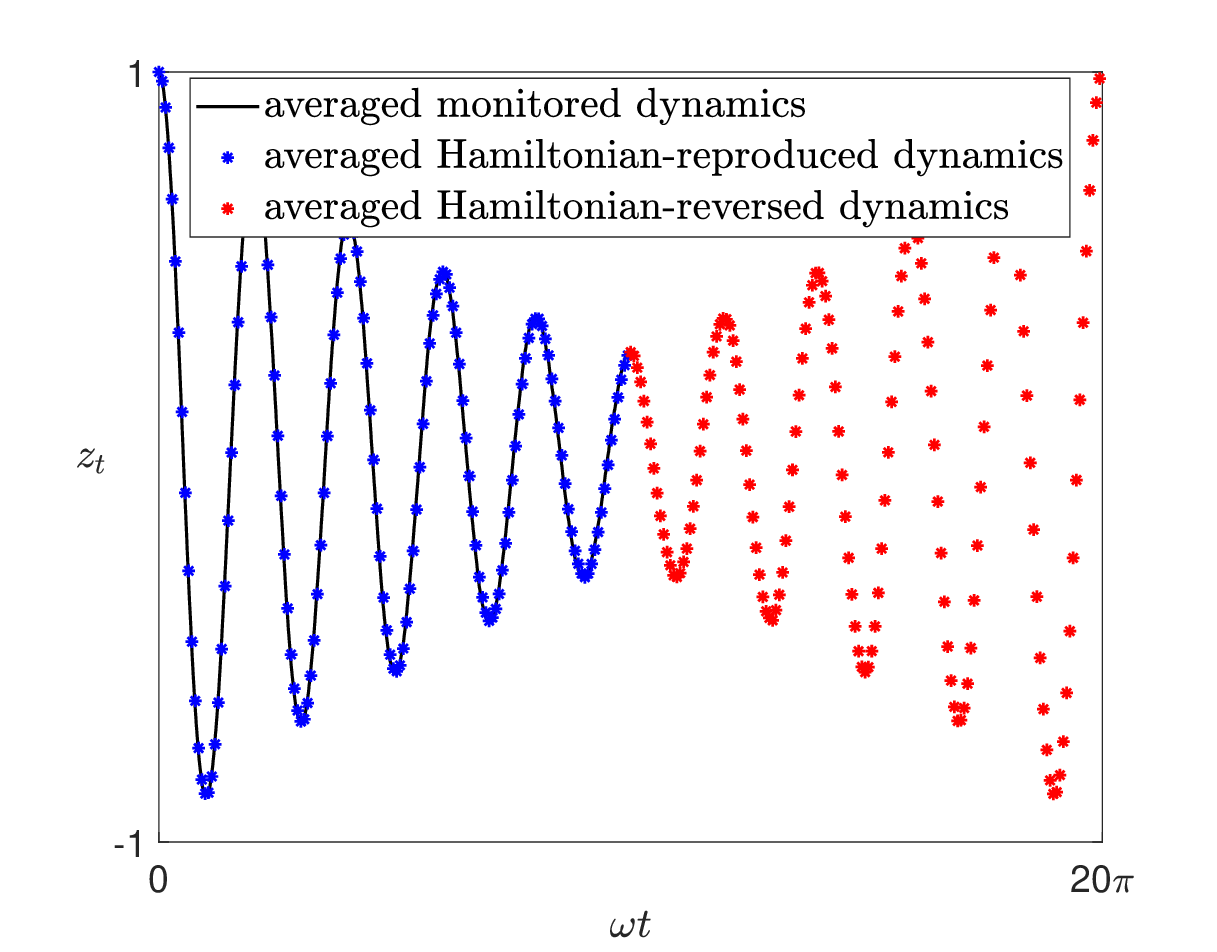} \\
\vspace{7pt}
{\footnotesize (b) Entropy of forward and backward-in-time emulated dynamics}
\includegraphics[width=0.48\textwidth]{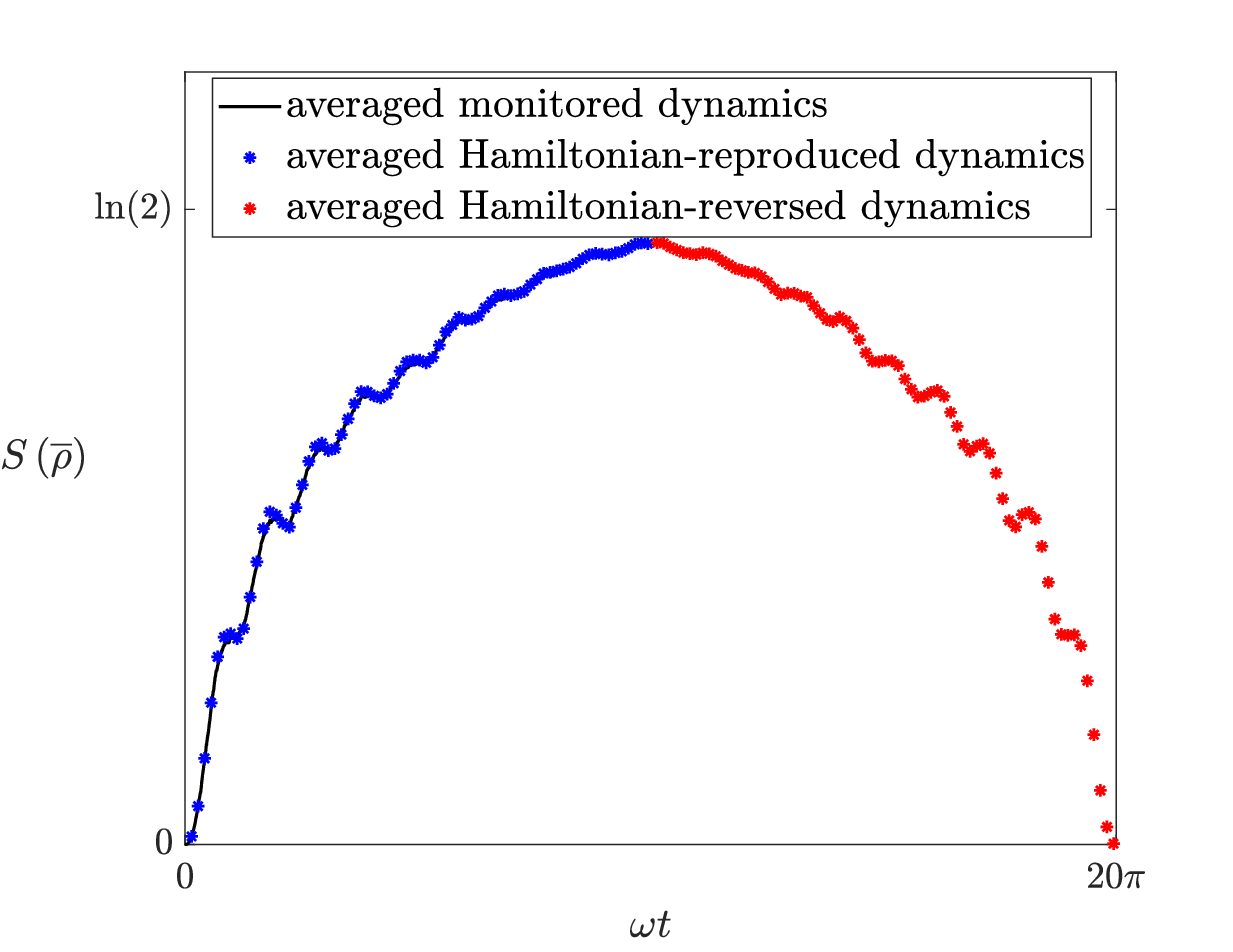} 
\caption{\label{fig:fig3}
[Plot (a)]
The black continuous curve denotes the averaged evolution over many realizations of a monitored qubit's $z_t = \tr{\rho_t \sigma_z}$ coordinate as a function of time, with $\omega \tau = 2\pi$ and $\tau/dt = 10^3$. The blue dots represent the evolution obtained by averaging over trajectories reproduced by $\Hmeas^\xi$. That is, the blue dots are a Hamiltonian emulation of the dynamics of the open system. The red dots are the average of the time-reverse trajectories, driven realizations of $-H -\Hmeas^\xi$ from the final to the initial states of the forward processes. This illustrates the Hamiltonian emulation of the backward-in-time dynamics of an open quantum system. [Plot (b)] The von Neumann $S(\overline{\rho_{\xi,t}})$ entropy of the averaged state $\overline{\rho_{\xi,t}}$ increases in the emulation of the open dynamics, but decreases in the emulation of the backward-in-time dynamics.
}
\end{figure}

To accomplish such a task, one could repeat runs of a continuous measurement protocol to record realizations $\xi$ of the output $\rvec^\xi$. For each output $\rvec^\xi$, classically simulating the state's update rule [Eqs.~\eqref{eq:Kraus} and~\eqref{eq-forward}]  yields a known state trajectory of pure states, from which one can calculate $\Hmeas^\xi$ via Eq.~\eqref{eq:Hmeas}. 
Given each known $\Hmeas^\xi$, one would drive the system with the Hamiltonian $H+\Hmeas^\xi$ in an experiment. 
Averaging this process over sufficiently many runs emulates the dynamics of an open system in a Hamiltonian way.
Alternatively, one could simulate measurement processes on a classical computer to find (simulated) measurement records and their trajectories, from which instances of $\Hmeas^\xi$ could be calculated. Driving a system with such $\Hmeas^\xi$, calculated from classical simulations, and averaging would also emulate open quantum dynamics.

Given the previous protocols, it is also straightforward to simulate time-reversed dynamics. 
At each time $t$ and for each realization of $\Hmeas^\xi \equiv \Hmeas^\xi(t)$, evolving with $H + \Hmeas^\xi(t)$ reproduces the effect of monitoring, taking the system from the pure state $\rho_{\xi,t}$ to the pure state $\rho_{\xi,t+dt}$. Thus, evolving with $-H - \Hmeas^\xi(t)$ simulates the effect of time reversal at time $t$, driving the system from the pure state $\rho_{\xi,t+dt}$ back in time to the pure state $\rho_{\xi,t}$. Starting from the final state $\rho_{\xi,T}$ of a stochastic trajectory and evolving with $-H - \Hmeas^\xi(T-t)$ emulates a backward trajectory, from $\rho_{\xi,T}$ to $\rho_0$. Finally, averaging over realizations of such back-in-time trajectories yields the time-reversed dynamics of an open quantum system. We illustrate the emulated time-reversed dynamics of a qubit in Fig.~\ref{fig:fig3}. 
(Our method is complementary to the open dynamics simulated by engineered noise~\cite{PhysRevLett.118.140403}. Note, also, that knowledge of the full trajectory is needed, so our method differs from works that aim to drive a system back in time to an unknown initial state~\cite{Aharonov1990SuperpositionsOT, Miguel1, Miguel2, Miguel3, PRLReversing}.)

\subsection{A continuous quantum measurement engine}
\label{sec:engine}
\begin{figure*}[!htbp]
{\footnotesize (a) Internal energy of the continuous measurement engine }  \qquad\quad\quad   
{\footnotesize (b) Energy output of the continuous measurement engine }
\includegraphics[width=0.47\textwidth]{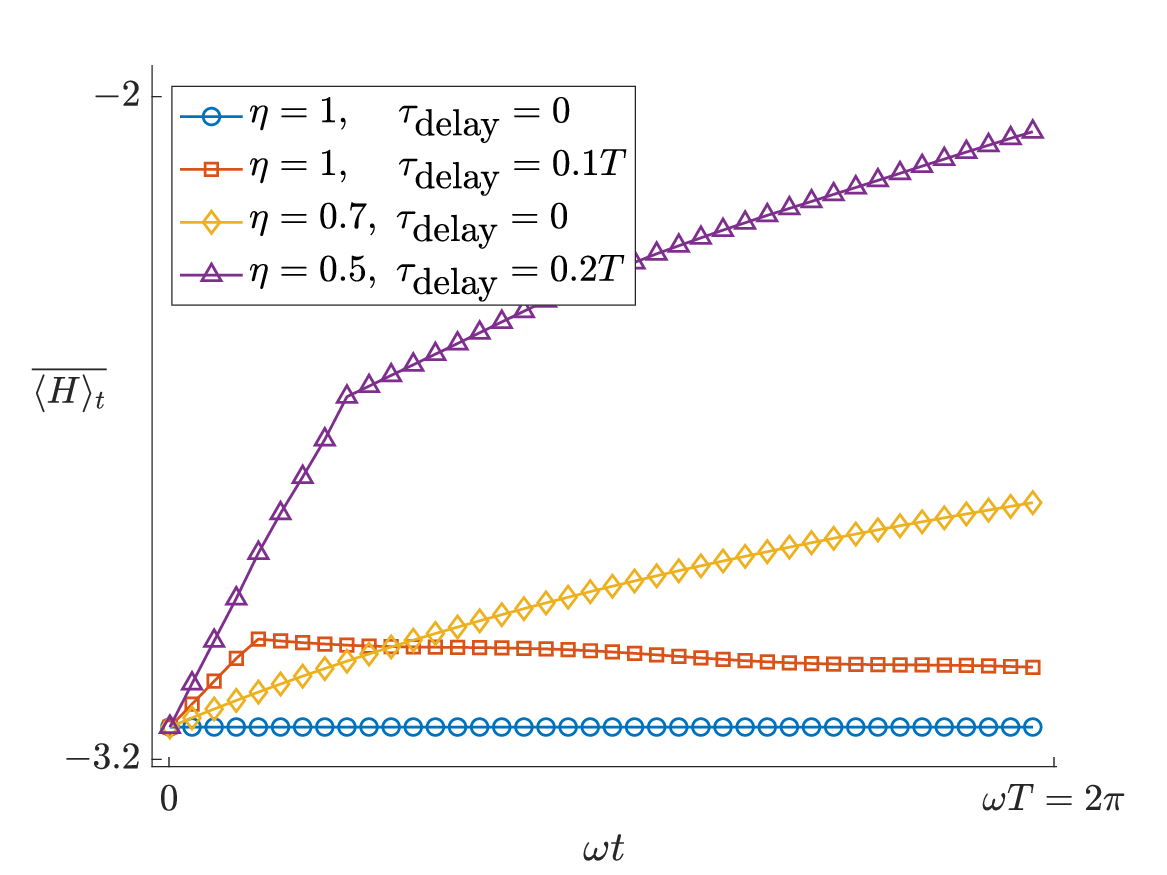} 
 \vspace{21pt}
\includegraphics[width=0.47\textwidth]{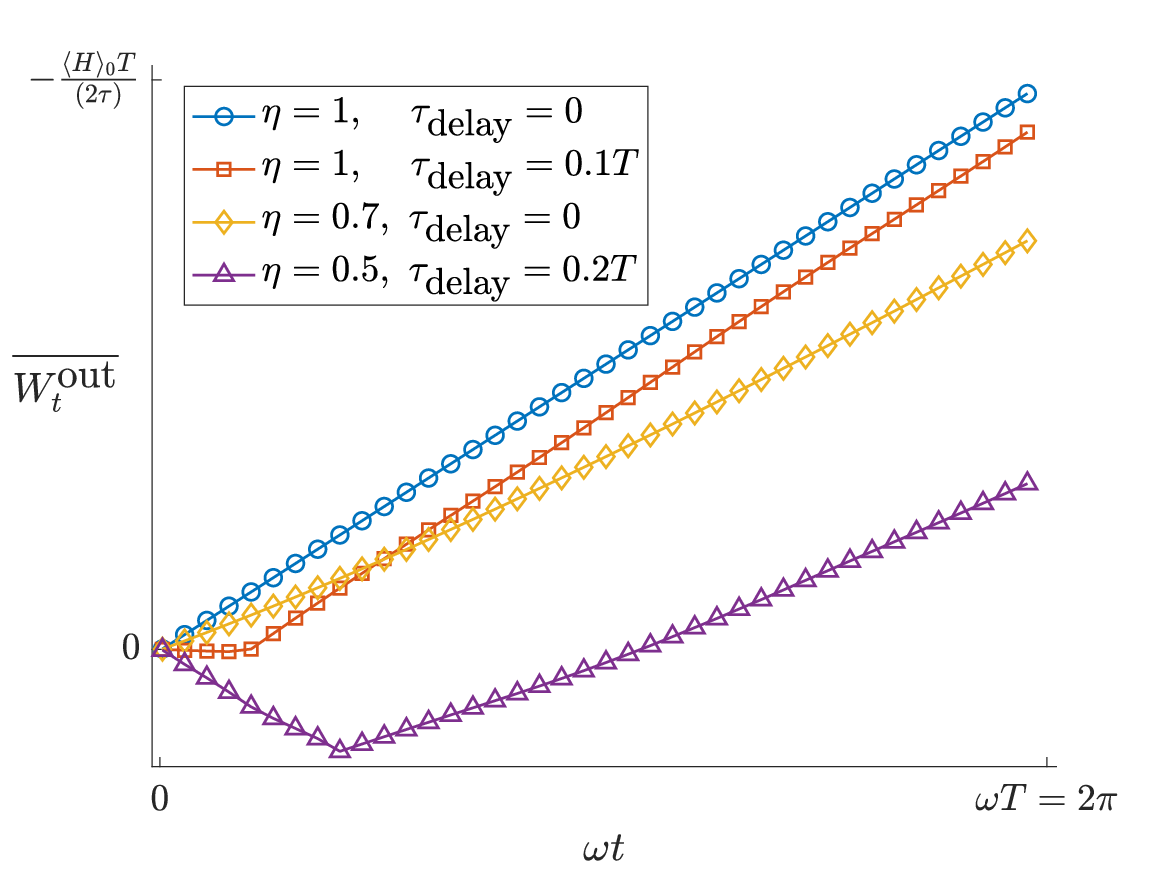} 
\vspace{-21pt}
\caption{\label{fig:fig4}
We consider the two-level continuous measurement engine introduced in Sec.~\ref{sec:engine}. 
Without feedback, the trajectory-averaged energy of the qubit increases according to $\overline{\langle H \rangle_t} = e^{-t/(2\tau)} \langle H \rangle_0$ [see paragraph before Eq.~\eqref{eq:EnergyMeas} and note that $\langle H \rangle_0 < 0$ for the system to act as an engine]. Feedback with $\Hfb$ and $\chron = -1$ slows down the rate at which energy increases. The energy that would have been pumped into the system is thus extracted by the agent performing feedback.
Figures (a) and (b) show the system's internal energy and the corresponding engine's output for different values of the measurement efficiency $\eta$ and the feedback delay $\tau_{\textnormal{delay}}$. The engine's output is calculated by subtracting the system's energy from the energy the system would have if no feedback took place.  
Under ideal measurement and feedback ($\eta = 1$ and $\tau_{\textnormal{delay}} = 0$, blue curves with circles), the system is pinned to its initial state. In such case, all the energy that would have been pumped into the system is extracted.
Under feedback delay and efficient measurements ($\eta = 1$ and $\tau_{\textnormal{delay}} = 0.1T$, orange curves with squares), the engine has null output until feedback starts acting. Without delay but inefficient measurements ($\eta = 0.7$ and $\tau_{\textnormal{delay}} = 0$, yellow curves with rhombi), the engine's work output diminishes. Finally, under inefficient measurements and feedback delay ($\eta = 0.5$ and $\tau_{\textnormal{delay}} = 0.2T$, purple curves with triangle), the engine has negative output for a period of time, but eventually reaches positive output. Note that the regimes considered are overly conservative. Experimental measurement efficiencies on superconducting qubits typically fall within the range $\eta_{\textnormal{meas}} \in [0.7,0,8]$~\cite{PhysRevLett.126.020502,PhysRevX.9.011004}, and feedback delay is typically orders of magnitude below qubit coherence times~\cite{PhysRevX.3.021008}. 
}
\end{figure*}

Measurements can pump energy into or draw energy out of a quantum system. To see this, note that the state $\overline{\rho}_t$ obtained by averaging over measurement outcomes satisfies the Lindblad master equation $d\overline{\rho}_t \nobreak=\nobreak -i[H,\overline{\rho}_t]dt \nobreak-\nobreak \tfrac{dt}{8\tau} [A,[A,\overline{\rho}_t]]$. The system's average energy satisfies $d\overline{\langle H \rangle}_t \nobreak=\nobreak - \tfrac{dt}{8\tau} \tr{\overline{\rho}_t [[H,A],A]}$, so it increases if, for instance, the initial state is the eigenstate of $H$ with the minimum energy and $[H,A]\neq 0$. For a qubit with $H = \omega \sigma_y/2$ and $A = \sigma_z$, the average energy follows $d\overline{\langle H \rangle}_t \nobreak=\nobreak - \overline{\langle H \rangle}_t dt/(2\tau)$.

More generally, the system's internal energy change due to measurements is
\begin{align}
\label{eq:EnergyMeas}
  d \langle H \rangle_t 
  = \frac{r_t dt}{\tau} \cov_t(A,H),  
\end{align}
where $\cov_t \left( A, H \right) \coloneqq \tr{\rho_t \{A,H\}}/2 - \tr{\rho_t A} \tr{\rho_t A}$ is the covariance between $A$ and $H$.
Equation~\eqref{eq:EnergyMeas} can be derived from Eq.~\eqref{eq:Strato} or, equivalently, from Schr\"odinger's equation with the measurement Hamiltonian $\Hmeas$ in Eq.~\eqref{eq:Hmeas} (we assume $A^2 = \id$ for simplicity). Upon averaging over measurement realizations, one has $\overline{d\langle H \rangle_t} \nobreak = \nobreak \frac{dt}{\tau} \overline{\langle A \rangle_t \cov_t(A,H)}$. Thus, the sign of $\langle A \rangle_t \cov_t(A,H)$ determines the average direction of energy flow.

The feedback process changes the system's energy, too, by performing work~\cite{Strasberg_2021,manzano2022quantum}. We show in Appendix~\ref{sec-app:engine} that the rate at which the feedback-driven process does thermodynamic work is given by
\begin{align}
\label{eq:work}
    \delta W_t \coloneqq \tr{\rho_t d \Hfb} = \chron \frac{r_t dt}{\tau} \cov_t \left( A, H \right),
\end{align}
where we adopt the convention of positive work when there is a cost to the process, and negative if energy is extracted from the process.

The work $\delta W_t$ performed by the feedback agent is thus proportional to the energy change due to measurements, Eq.~\eqref{eq:EnergyMeas}. However, its sign and magnitude depend on $\chron$. The arrow-reshaping parameter $\chron $ can then be interpreted as the quotient between the work performed by the feedback agent and the energy flow from measurements. If feedback enhances the energy flow due to measurements, then $\chron > 0$, which stretches the arrow of time. When feedback goes against the energy flow from measurements, then $\chron < 0$, which shrinks or even inverts the arrow of time (see Sec.~\ref{sec:Reshaping}).

The previous discussion gives a recipe to design measurement engines. When $\delta W_t < 0$, the feedback extracts energy from the system, drawing it from the monitoring. Given knowledge of the system's state and readout $r_t$, choosing $\textnormal{sign}(\chron) = - \textnormal{sign}(d\langle H\rangle_t) = - \textnormal{sign}\big( r_t \cov_t(A,H) \big)$ ensures that the feedback process acts as an engine that draws energy from measurements.

A greedy engine design could, for instance, alternate the sign of $\chron$ to ensure work extraction from each measurement readout and augment it by a large $|\chron|$. Such an engine has the disadvantage, though, of requiring knowledge of the system's rapidly changing state to construct the feedback Hamiltonian $\Hfb$ in Eq.~\eqref{eq:Hfb}. This would not only incur a computational cost, but also a thermodynamic cost from Landauer's principle~\cite{Landauer, reeb2014improved, faist2015minimal, PhysRevX.8.021011}. Instead, we propose an operating regime where the engine does not suffer from such a shortcoming.
 
Consider the system initialized in a known state $\rho_0$ with $\langle A \rangle_{0} \cov_0(A,H) > 0$, so that, on average, measurements are pumping energy into the system [see the discussion after Eq.~\eqref{eq:EnergyMeas}]. If we set $\chron = -1$, then $H_{\textnormal{fback}}^{\chron = -1} = + i r_t/(2\tau)[\rho_0,A]$ functions as an engine, with $\delta W_t < 0$, while also undoing the effect of measurements. The system is thus pinned in the initial state $\rho_0$ and feedback continuously extracts energy. Crucially, the feedback Hamiltonian's only time dependence is via the output $r_t$, readily available from the measurement device. Thus, the correct feedback can be applied without needing to simulate the time-evolution of the system. 

For a qubit with $H = \omega \sigma_y/2$ where $A = \sigma_z$ is continuously monitored, we show in Appendix~\ref{sec-app:engine} that the average work $\overline{W^\textnormal{out}_T}$ extracted by the feedback over a time $T$ for $\chron = -1$ is
\begin{align}
\label{eq:workrate}
   \overline{W^\textnormal{out}_T}  = -\frac{\langle H \rangle_0}{2\tau} T.
\end{align}
[Note a global minus sign in Eq.~\eqref{eq:workrate} when focusing on the work extracted by the feedback agent, as opposed to the work done on the system.]

Since $\Hfb$ counteracts measurements at $\chron = -1$, the qubit's energy $\langle H \rangle_0$ is constant. Thus, for $\langle H \rangle < 0$, the system functions as an engine that draws power from measurements. The power $\overline{W^\textnormal{out}_T}/T$ from Eq.~\eqref{eq:workrate} coincides with the rate at which the system's energy changes by monitoring [see the paragraph before Eq.~\eqref{eq:EnergyMeas}], so $H_{\textnormal{fback}}^{\chron = -1}$ extracts all the energy pumped by measurements. 

References~\cite{MEnginePRL2017, MEnginePRL2018} introduced pulsed feedback to drive measurement engines, and Ref.~\cite{PhysRevA.106.042221} studied the interplay between the feedback agent's perceived arrow of time and a measurement engine's output. Continuous feedback was considered to charge quantum batteries~\cite{Mitchison2021chargingquantum}, cool quantum systems~\cite{bhandari2023measurement,de2024continuous, elouard2025revealing}, and to drive engines~\cite{PhysRevE.105.044137, dassonneville}. Unlike the models in the aforementioned references, our feedback process fixes the engine in a steady state $\rho_0$. $H_{\textnormal{fback}}^{\chron = -1} = +i \frac{r_t}{2\tau}  \left[\rho_0,A \right]$, carefully counterbalances measurements. In the process, the energy pumped into the system by the monitoring process is extracted by the feedback procedure. This happens in the regime where feedback is shrinking time's arrow.

Under experimentally realistic conditions, measurement devices only capture partial information. Measurement losses can be modeled by an efficiency parameter $0 \leq \eta \leq 1$, which quantifies the fraction of measurement outcomes captured~\cite{JacobsIntro2006}. Moreover, feedback processes incur a delay time $\tau_{\textnormal{delay}}$ between the time $t$ when a measurement outcome is read and the time $t+\tau_{\textnormal{delay}}$ when the feedback Hamiltonian acts. The ideal feedback-measurement processes we have described so far assumed $\eta = 1$ and $\tau_{\textnormal{delay}} = 0$.

In Fig.~\ref{fig:fig4}, we simulate the output of a continuous quantum measurement engine under experimentally realistic measurement efficiencies and feedback delays. We show that, despite such non-idealities, the feedback process continuously extracts energy from measurements, albeit at a reduced output.

\section{Discussion}

Maxwell's demon exploits knowledge of a classical system's state to generate thermodynamically anomalous processes, such as decreasing the entropy of an otherwise isolated system. The thermodynamic arrow of time appears to flow backward in such situations. Our modern version of Maxwell's demon exploits knowledge of a quantum system's state and measurement outcomes to drive similarly anomalous processes, altering an observer's perception of the quantum arrow of time.

In our framework, the agent relies on the Hamiltonian $\Hmeas$ [Eq.~\eqref{eq:Hmeas}] which, given knowledge of a system's initial state and a sequence of measurement outcomes, reproduces the stochastic trajectory of a monitored quantum system. Leveraging the explicit expression for $\Hmeas$ allowed the design of feedback protocols that can: 
\begin{itemize}
    \item[(i)] generate stochastic trajectories more consistent with a backward arrow of time than a forward one,
    \item[(ii)] simulate the forward and time-reversed dynamics of an open quantum system, and
    \item[(iii)] drive a measurement engine that continuously draws all the energy pumped into a system by measurements.
\end{itemize}

We have found intricate connections between the operating regimes of measurement engines, the flow of energy from measurements, and the way in which feedback affects the perceived arrow of time. Time reversal is not always needed to extract work. If measurements are, on average, taking energy out of the system, the feedback mechanism that stretches time's arrow also extracts work. However, the system's energy will rapidly increase, possibly saturating for bounded Hamiltonians. Moreover, in this operating regime the engine's state is changing, so the feedback mechanism requires a computationally costly simulation to find the corresponding feedback Hamiltonian.

In contrast, we have found a stable regime where shrinking the arrow of time ($\chron = -1$) pins the engine in a fixed state. In such a regime, the engine continuously outputs work at a steady rate. 

Demonstrating (i) and (ii) requires calculating the trajectory of a state from the measurement output to reconstruct the trajectory-dependent $\Hmeas$. However, the state of the engine remains constant in application (iii), so demonstrating it only requires knowledge of the measurement output to implement $H_{\textnormal{fback}}^{\chron = -1} = -\Hmeas$. Perhaps the most natural next step is to experimentally demonstrate the use of $\Hmeas$ for quantum feedback control, e.g., in superconducting qubits; a platform that allows for rapid feedback and high detection efficiencies~\cite{Minev_2019} and in which quantum versions of Maxwell's demon have been implemented~\cite{HuardMaxwellDemon, MurchMaxwellDemon}.

\section*{Acknowledgements}
We thank Ian Spielman, Philippe Lewalle, and Aar\'on Villanueva for discussions about continuous feedback control and stochastic calculus. 
This material is based upon work supported by the U.S. Department of Energy, Office of Science, Accelerated Research in Quantum Computing, Fundamental Algorithmic Research toward Quantum Utility (FAR-Qu) and Fundamental Algorithmic Research in Quantum Computing (FAR-QC). We also acknowledge support from the Beyond Moore’s Law project of the Advanced Simulation and Computing Program at LANL managed by Triad National Security, LLC, for the National Nuclear Security Administration of the U.S. DOE under contract 89233218CNA000001. A.V.G.~also acknowledges support by the NSF QLCI (award No.~OMA-2120757), NSF STAQ program, DoE ASCR Quantum Testbed Pathfinder program (awards No.~DE-SC0019040 and No.~DE-SC0024220), AFOSR MURI, ARL (W911NF-24-2-0107), DARPA SAVaNT ADVENT, and NQVL:QSTD:Pilot:FTL. A.V.G.~also acknowledges support from the U.S.~Department of Energy, Office of Science, National Quantum Information Science Research Centers, Quantum Systems Accelerator.

\bibliography{references}

@article{hastings2022lieb,
  title={On {Lieb-Robinson} bounds for the double bracket flow},
  author={Hastings, Matthew B},
  journal={arXiv preprint arXiv:2201.07141},
  doi = {10.48550/arXiv.2201.07141},
  year={2022}
}

@article{acin2016certified,
  title={Certified randomness in quantum physics},
  author={Ac{\'\i}n, Antonio and Masanes, Lluis},
  journal={Nature},
  volume={540},
  doi = {10.1038/nature20119},
  number={7632},
  pages={213--219},
  year={2016},
  publisher={Nature Publishing Group UK London}
}

@article{pironio2010random,
  title={Random numbers certified by {Bell}’s theorem},
doi = {10.1038/nature09008},
  author={Pironio, Stefano and Ac{\'\i}n, Antonio and Massar, Serge and de La Giroday, A Boyer and Matsukevich, Dzmitry N and Maunz, Peter and Olmschenk, Steven and Hayes, David and Luo, Lefroy and Manning, T Andrew and others},
  journal={Nature},
  volume={464},
  number={7291},
  pages={1021--1024},
  year={2010},
  publisher={Nature Publishing Group UK London}
}

@article{collin2005verification,
  title={{Verification of the Crooks fluctuation theorem and recovery of RNA folding free energies}},
  author={Collin, Delphine and Ritort, Felix and Jarzynski, Christopher and Smith, Steven B and Tinoco Jr, Ignacio and Bustamante, Carlos},
  journal={Nature},
  volume={437},
  number={7056},
  pages={231--234},
  doi = {10.1038/nature04061},
  year={2005},
  publisher={Nature Publishing Group UK London}
}

@article{roldan2021quantifying,
  title={Quantifying entropy production in active fluctuations of the hair-cell bundle from time irreversibility and uncertainty relations},
  author={Rold{\'a}n, {\'E}dgar and Barral, J{\'e}r{\'e}mie and Martin, Pascal and Parrondo, Juan MR and J{\"u}licher, Frank},
  journal={New J. Phys.},
  volume={23},
  number={8},
  pages={083013},
  year={2021},
  doi = {10.1088/1367-2630/ac0f18},
  publisher={IOP Publishing}
}

@book{frank2022physics,
  title={The physics and mathematics of {Elliott Lieb}},
  author={Frank, Rupert L and Laptev, Ari and Lewin, Mathieu and Seiringer, Robert},
  year={2022},
  doi = {10.4171/90},
  isbn = {978-3-98547-519-3},
  publisher={EMS Press}
}

@article{Gluza2024doublebracket,
  doi = {10.22331/q-2024-04-09-1316},
  url = {https://doi.org/10.22331/q-2024-04-09-1316},
  title = {Double-bracket quantum algorithms for diagonalization},
  author = {Gluza, Marek},
  journal = {{Quantum}},
  issn = {2521-327X},
  publisher = {{Verein zur F{\"{o}}rderung des Open Access Publizierens in den Quantenwissenschaften}},
  volume = {8},
  pages = {1316},
  month = apr,
  year = {2024}
}

@article{BROCKETT199179,
title = {Dynamical systems that sort lists, diagonalize matrices, and solve linear programming problems},
journal = {Linear Algebra Appl.},
volume = {146},
pages = {79-91},
year = {1991},
issn = {0024-3795},
doi = {10.1016/0024-3795(91)90021-N},
url = {https://www.sciencedirect.com/science/article/pii/002437959190021N},
author = {R.W. Brockett}
}

@article{lees2012observation,
  title={{Observation of time-reversal violation in the B 0 meson system}},
  author={Lees, J.P. and Poireau, V and Tisserand, V and Garra Tico, J and Grauges, E and Palano, A and Eigen, G and Stugu, B and Brown, David Nathan and Kerth, LT and others},
  journal={Phys. Rev. Lett.},
  volume={109},
  doi={10.1103/PhysRevLett.109.211801},
  number={21},
  pages={211801},
  year={2012},
  publisher={APS}
}

@article{PhysRevX.9.011004,
  title = {High-Efficiency Measurement of an Artificial Atom Embedded in a Parametric Amplifier},
  author = {Eddins, A. and Kreikebaum, J. M. and Toyli, D. M. and Levenson-Falk, E. M. and Dove, A. and Livingston, W. P. and Levitan, B. A. and Govia, L. C. G. and Clerk, A. A. and Siddiqi, I.},
  journal = {Phys. Rev. X},
  volume = {9},
  issue = {1},
  pages = {011004},
  numpages = {11},
  year = {2019},
  month = {Jan},
  publisher = {American Physical Society},
  doi = {10.1103/PhysRevX.9.011004},
  url = {https://link.aps.org/doi/10.1103/PhysRevX.9.011004}
}

@article{PhysRevLett.126.020502,
  title = {Efficient Qubit Measurement with a Nonreciprocal Microwave Amplifier},
  author = {Lecocq, F. and Ranzani, L. and Peterson, G. A. and Cicak, K. and Jin, X. Y. and Simmonds, R. W. and Teufel, J. D. and Aumentado, J.},
  journal = {Phys. Rev. Lett.},
  volume = {126},
  issue = {2},
  pages = {020502},
  numpages = {5},
  year = {2021},
  month = {Jan},
  publisher = {American Physical Society},
  doi = {10.1103/PhysRevLett.126.020502},
  url = {https://link.aps.org/doi/10.1103/PhysRevLett.126.020502}
}

@article{GarrahanPRB2025,
  title = {Efficient post-selection in light cone correlations of monitored quantum circuits},
  author = {Li, Jimin and Jack, Robert L. and Bertini, Bruno and Garrahan, Juan P.},
  journal = {Phys. Rev. B},
  volume = {111},
  issue = {2},
  pages = {024309},
  numpages = {16},
  year = {2025},
  month = {Jan},
  publisher = {American Physical Society},
  doi = {10.1103/PhysRevB.111.024309},
  url = {https://link.aps.org/doi/10.1103/PhysRevB.111.024309}
}

@article{alipourQuantum2020,
  title={Shortcuts to adiabaticity in driven open quantum systems: Balanced gain and loss and non-Markovian evolution},
  author={Alipour, Sahar and Chenu, Aurelia and Rezakhani, Ali T and del Campo, Adolfo},
  journal={Quantum},
  volume={4},
  pages={336},
  year={2020},
  doi={10.22331/q-2020-09-28-336},
  publisher={Verein zur F{\"o}rderung des Open Access Publizierens in den Quantenwissenschaften}
}

@article{alipourPRA2022,
  title = {Entropy-based formulation of thermodynamics in arbitrary quantum evolution},
  author = {Alipour, S. and Rezakhani, A. T. and Chenu, A. and del Campo, A. and Ala-Nissila, T.},
  journal = {Phys. Rev. A},
  volume = {105},
  issue = {4},
  pages = {L040201},
  numpages = {6},
  year = {2022},
  month = {Apr},
  publisher = {American Physical Society},
  doi = {10.1103/PhysRevA.105.L040201},
  url = {https://link.aps.org/doi/10.1103/PhysRevA.105.L040201}
}

@article{
HuardMaxwellDemon,
author = {Nathanaël Cottet  and Sébastien Jezouin  and Landry Bretheau  and Philippe Campagne-Ibarcq  and Quentin Ficheux  and Janet Anders  and Alexia Auffèves  and Rémi Azouit  and Pierre Rouchon  and Benjamin Huard },
title = {Observing a quantum {Maxwell} demon at work},
journal = {PNAS},
volume = {114},
number = {29},
pages = {7561-7564},
year = {2017},
doi = {10.1073/pnas.1704827114}
}

@misc{hu2023describingwavefunctioncollapse,
      title={Describing the Wave Function Collapse Process with a State-dependent Hamiltonian}, 
      author={Le Hu and Andrew N. Jordan},
      year={2023},
      eprint={2301.09274},
      archivePrefix={arXiv},
      primaryClass={quant-ph} 
}

@article{bhandari2023measurement,
  title={Measurement-based quantum thermal machines with feedback control},
  author={Bhandari, Bibek and Czupryniak, Robert and Erdman, Paolo Andrea and Jordan, Andrew N},
  journal={Entropy},
  volume={25},
  number={2},
  pages={204},
  doi={1099-4300/25/2/204},
  year={2023},
  publisher={MDPI}
}

@ARTICLE{Landauer,
  author={Landauer, R.},
  journal={IBM J. Res. Dev.}, 
  title={Irreversibility and Heat Generation in the Computing Process}, 
  year={1961},
  volume={5},
  number={3},
  pages={183-191},
  doi={10.1147/rd.53.0183}}

@article{reeb2014improved,
  title={An improved Landauer principle with finite-size corrections},
  author={Reeb, David and Wolf, Michael M},
  journal={New J. Phys.},
  doi = {10.1088/1367-2630/16/10/103011},
  volume={16},
  number={10},
  pages={103011},
  year={2014},
  publisher={IOP Publishing}
}

@article{faist2015minimal,
  title={The minimal work cost of information processing},
  author={Faist, Philippe and Dupuis, Fr{\'e}d{\'e}ric and Oppenheim, Jonathan and Renner, Renato},
  journal={Nat. Commun.},
  volume={6},
  number={1},
  doi = {10.1038/ncomms8669},
  pages={7669},
  year={2015},
  publisher={Nature Publishing Group UK London}
}

@article{PhysRevX.8.021011,
  title = {Fundamental Work Cost of Quantum Processes},
  author = {Faist, Philippe and Renner, Renato},
  journal = {Phys. Rev. X},
  volume = {8},
  issue = {2},
  pages = {021011},
  numpages = {19},
  year = {2018},
  month = {Apr},
  publisher = {American Physical Society},
  doi = {10.1103/PhysRevX.8.021011},
  url = {https://link.aps.org/doi/10.1103/PhysRevX.8.021011}
}

@article{PhysRevLett.118.140403,
  title = {Quantum Simulation of Generic Many-Body Open System Dynamics Using Classical Noise},
  author = {Chenu, A. and Beau, M. and Cao, J. and del Campo, A.},
  journal = {Phys. Rev. Lett.},
  volume = {118},
  issue = {14},
  pages = {140403},
  numpages = {6},
  year = {2017},
  month = {Apr},
  publisher = {American Physical Society},
  doi = {10.1103/PhysRevLett.118.140403},
  url = {https://link.aps.org/doi/10.1103/PhysRevLett.118.140403}
}

@article{GisinStrato,
doi = {10.1088/1355-5111/8/1/018},
url = {https://dx.doi.org/10.1088/1355-5111/8/1/018},
year = {1996},
month = {feb},
publisher = {},
volume = {8},
number = {1},
pages = {255},
author = {M. Rigo and N. Gisin},
title = {Unravellings of the master equation and the emergence of a  classical world},
journal = {Quantum Semiclass. Opt.}
}

@article{PhysRevX.3.021008,
  title = {Persistent Control of a Superconducting Qubit by Stroboscopic Measurement Feedback},
  author = {Campagne-Ibarcq, P. and Flurin, E. and Roch, N. and Darson, D. and Morfin, P. and Mirrahimi, M. and Devoret, M. H. and Mallet, F. and Huard, B.},
  journal = {Phys. Rev. X},
  volume = {3},
  issue = {2},
  pages = {021008},
  numpages = {7},
  year = {2013},
  month = {May},
  publisher = {American Physical Society},
  doi = {10.1103/PhysRevX.3.021008},
  url = {https://link.aps.org/doi/10.1103/PhysRevX.3.021008}
}

@article{Aharonov1990SuperpositionsOT,
  title={Superpositions of time evolutions of a quantum system and a quantum time-translation machine.},
  author={Yakir Aharonov and Yakir Aharonov and Jeeva S. Anandan and Jeeva S. Anandan and Sandu Popescu and Lev Vaidman and Lev Vaidman},
  journal={Phys. Rev. Lett.},
  year={1990},
  volume={64 25},
  pages={
          2965-2968
        },
  doi = {10.1103/PHYSREVLETT.64.2965}
}

@article{de2024continuous,
  title = {Continuous feedback protocols for cooling and trapping a quantum harmonic oscillator},
  author = {De Sousa, Guilherme and Bakhshinezhad, Pharnam and Annby-Andersson, Bj\"orn and Samuelsson, Peter and Potts, Patrick P. and Jarzynski, Christopher},
  journal = {Phys. Rev. E},
  volume = {111},
  issue = {1},
  pages = {014152},
  numpages = {17},
  year = {2025},
  month = {Jan},
  publisher = {American Physical Society},
  doi = {10.1103/PhysRevE.111.014152},
  url = {https://link.aps.org/doi/10.1103/PhysRevE.111.014152}
}

@article{Strasberg_2021,
  title = {First and Second Law of Quantum Thermodynamics: A Consistent Derivation Based on a Microscopic Definition of Entropy},
  author = {Strasberg, Philipp and Winter, Andreas},
  journal = {PRX Quantum},
  volume = {2},
  issue = {3},
  pages = {030202},
  numpages = {26},
  year = {2021},
  month = {Aug},
  publisher = {American Physical Society},
  doi = {10.1103/PRXQuantum.2.030202},
  url = {https://link.aps.org/doi/10.1103/PRXQuantum.2.030202}
}

@article{LordKelvin,
author={Thomson, William},
title={Kinetic Theory of the Dissipation of Energy},
journal={Nature},
year={1874},
month={Apr},
day={01},
volume={9},
number={232},
pages={441-444},
issn={1476-4687},
doi={10.1038/009441c0}
}

@book{Maxwell,
 place={Cambridge}, 
 series={Cambridge Library Collection - Physical  Sciences}, 
 title={Theory of Heat}, publisher={Cambridge University Press}, 
 author={Maxwell, James Clerk},
  year={2011}, 
  doi ={10.1017/CBO9781139057943},
  collection={Cambridge Library Collection - Physical  Sciences}
  }

@book{gardinerstochastic,
	author = {Gardiner, Crispin W. and Zoller, P.},
	title = {Quantum noise: a handbook of markovian and non-markovian quantum stochastic methods with applications to quantum optics},
	publisher = {Springer},
	year = {2004},
	series = {Springer series in synergetics},
	address = {New York},
	edition = {3rd ed.}
}

@article{jackson2023simultaneous,
  title={Simultaneous Measurements of Noncommuting Observables: Positive Transformations and Instrumental {Lie} Groups},
  author={Jackson, Christopher S and Caves, Carlton M},
  journal={Entropy},
  volume={25},
  number={9},
  pages={1254},
  doi={10.3390/e25091254},
  year={2023},
  publisher={MDPI}
}

@inbook{Penrose:1980ge,
    author = "Penrose, R.",
    title = "{Singularities and Time Asymmetry}",
    booktitle = "{General Relativity}: {An Einstein Centenary Survey}",
    pages = "581--638",
    isbn = "978-0-521-29928-2",
    publisher = "Univ. Pr.",
    address = "Cambridge, UK",
    year = "1980"
}

@article{Jarzynski,
author = {Jarzynski, Christopher},
title = {Equalities and Inequalities: Irreversibility and the Second Law of Thermodynamics at the Nanoscale},
journal = {Annu. Rev. Condens. Matter Phys.},
volume = {2},
number = {1},
pages = {329-351},
year = {2011},
doi = {10.1146/annurev-conmatphys-062910-140506}
}

@article{prigogine2000arrow,
  title={The arrow of time},
  author={Prigogine, I},
  journal={The Chaotic Universe},
  pages={1-15},
  year={2000},
  doi = {10.1142/4374},
  publisher={World Scientific Singapore}
}

@article{MurchMaxwellDemon,
  title = {Information Gain and Loss for a Quantum {Maxwell's} Demon},
  author = {Naghiloo, M. and Alonso, J. J. and Romito, A. and Lutz, E. and Murch, K. W.},
  journal = {Phys. Rev. Lett.},
  volume = {121},
  issue = {3},
  pages = {030604},
  numpages = {6},
  year = {2018},
  month = {Jul},
  publisher = {American Physical Society},
  doi = {10.1103/PhysRevLett.121.030604},
  url = {https://link.aps.org/doi/10.1103/PhysRevLett.121.030604}
}

@article{Zurek_1998,
   title={Decoherence, Chaos, Quantum-Classical Correspondence, and
the Algorithmic Arrow of Time},
   volume={T76},
   ISSN={0031-8949},
   url={http://dx.doi.org/10.1238/Physica.Topical.076a00186},
   DOI={10.1238/physica.topical.076a00186},
   number={1},
   journal={Physica Scripta},
   publisher={IOP Publishing},
   author={Zurek, Wojciech H.},
   year={1998},
   pages={186} }

@article{superconducting2,
   title={Observing single quantum trajectories of a superconducting quantum bit},
   volume={502},
   ISSN={1476-4687},
   DOI={10.1038/nature12539},
   number={7470},
   journal={Nature},
   publisher={Springer Science and Business Media LLC},
   author={Murch, K. W. and Weber, S. J. and Macklin, C. and Siddiqi, I.},
   year={2013},
   month=oct, pages={211–214} }

@article{superconducting1,
author = {M. Hatridge  and S. Shankar  and M. Mirrahimi  and F. Schackert  and K. Geerlings  and T. Brecht  and K. M. Sliwa  and B. Abdo  and L. Frunzio  and S. M. Girvin  and R. J. Schoelkopf  and M. H. Devoret },
title = {Quantum Back-Action of an Individual Variable-Strength Measurement},
journal = {Science},
volume = {339},
number = {6116},
pages = {178-181},
year = {2013},
doi = {10.1126/science.1226897}
}

@ARTICLE{arrowlamb,
       author = {{Lamb}, Jeroen S.~W. and {Roberts}, John A.~G.},
        title = "{Time-reversal symmetry in dynamical systems: A survey}",
      journal = {Phys. D: Nonlinear Phenom.},
         year = 1998,
        month = jan,
       volume = {112},
       number = {1-2},
        pages = {1-39},
          doi = {10.1016/S0167-2789(97)00199-1},
       adsurl = {https://ui.adsabs.harvard.edu/abs/1998PhyD..112....1L}
}

@article{10.1162/netn_a_00300,
    author = {Deco, Gustavo and Sanz Perl, Yonatan and de la Fuente, Laura and Sitt, Jacobo D. and Yeo, B. T. Thomas and Tagliazucchi, Enzo and Kringelbach, Morten L.},
    title = "{The arrow of time of brain signals in cognition: Potential intriguing role of parts of the default mode network}",
    journal = {Netw. Neurosci.},
    volume = {7},
    number = {3},
    pages = {966-998},
    year = {2023},
    month = {10},
    issn = {2472-1751},
    doi = {10.1162/netn_a_00300}
}

@article{CrooksArrow,
  title = {Length of Time's Arrow},
  author = {Feng, Edward H. and Crooks, Gavin E.},
  journal = {Phys. Rev. Lett.},
  volume = {101},
  issue = {9},
  pages = {090602},
  numpages = {4},
  year = {2008},
  month = {Aug},
  publisher = {American Physical Society},
  doi = {10.1103/PhysRevLett.101.090602},
  url = {https://link.aps.org/doi/10.1103/PhysRevLett.101.090602}
}

@article{PhysRevResearch.5.033045,
  title = {Quantum state driving along arbitrary trajectories},
  author = {Hu, Le and Jordan, Andrew N.},
  journal = {Phys. Rev. Res.},
  volume = {5},
  issue = {3},
  pages = {033045},
  numpages = {7},
  year = {2023},
  month = {Jul},
  publisher = {American Physical Society},
  doi = {10.1103/PhysRevResearch.5.033045},
  url = {https://link.aps.org/doi/10.1103/PhysRevResearch.5.033045}
}

@article{RudolphPRE2010,
  title = {Entanglement and the thermodynamic arrow of time},
  author = {Jennings, David and Rudolph, Terry},
  journal = {Phys. Rev. E},
  volume = {81},
  issue = {6},
  pages = {061130},
  numpages = {9},
  year = {2010},
  month = {Jun},
  publisher = {American Physical Society},
  doi = {10.1103/PhysRevE.81.061130},
  url = {https://link.aps.org/doi/10.1103/PhysRevE.81.061130}
}

@article{ArrowIntitial1975,
 ISSN = {00368733, 19467087},
 URL = {http://www.jstor.org/stable/24949962},
 author = {David Layzer},
 journal = {Scientific American},
 number = {6},
 pages = {56--69},
 publisher = {Scientific American, a division of Nature America, Inc.},
 title = {THE ARROW OF TIME},
 urldate = {2022-10-05},
 volume = {233},
 year = {1975}
}

@article{wolpert2024memory,
AUTHOR = {Wolpert, David H. and Kipper, Jens},
TITLE = {Memory Systems, the Epistemic Arrow of Time, and the Second Law},
JOURNAL = {Entropy},
VOLUME = {26},
YEAR = {2024},
NUMBER = {2},
ARTICLE-NUMBER = {170},
ISSN = {1099-4300},
DOI = {10.3390/e26020170},
publisher={MDPI}
}

@book{BookZeh1989direction,
  title={The Direction of Time},
  author={Zeh, H-Dieter},
  year={1989},
  doi = {10.1007/978-3-540-68001-7},
  publisher={Springer}
}

@article{ellis2013arrow,
  title={The arrow of time and the nature of spacetime},
  author={Ellis, George Francis Rayner},
  journal={Stud. Hist. Philos. Sci. B - Stud. Hist. Philos. Mod. Phys.},
  volume={44},
  doi = {10.1016/j.shpsb.2013.06.002},
  number={3},
  pages={242--262},
  year={2013},
  publisher={Elsevier}
}

@book{BookHalliwell1996physical,
  title={Physical origins of time asymmetry},
  author={Halliwell, Jonathan J and P{\'e}rez-Mercader, Juan and Zurek, Wojciech Hubert},
  year={1996},
  publisher={Cambridge University Press}
}

@book{BookCarroll2010eternity,
  title={From eternity to here: the quest for the ultimate theory of time},
  author={Carroll, Sean},
  year={2010},
  publisher={Penguin}
}

@book{BookPrice1996time,
  title={Time's arrow \& Archimedes' point: new directions for the physics of time},
  author={Price, Huw},
  year={1996},
  publisher={Oxford University Press, USA}
}

@article{CosmoArrowHawking1985,
  title = {Arrow of time in cosmology},
  author = {Hawking, S. W.},
  journal = {Phys. Rev. D},
  volume = {32},
  issue = {10},
  pages = {2489--2495},
  numpages = {0},
  year = {1985},
  month = {Nov},
  publisher = {American Physical Society},
  doi = {10.1103/PhysRevD.32.2489},
  url = {https://link.aps.org/doi/10.1103/PhysRevD.32.2489}
}

@article{seif2021machine,
  title={Machine learning the thermodynamic arrow of time},
  author={Seif, Alireza and Hafezi, Mohammad and Jarzynski, Christopher},
  journal={Nat. Phys.},
  volume={17},
  doi = {10.1038/s41567-020-1018-2},
  number={1},
  pages={105--113},
  year={2021},
  publisher={Nature Publishing Group}
}

@ARTICLE{JacobsIntro2006,
   author = {{Jacobs}, K. and {Steck}, D.},
    title = "{A straightforward introduction to continuous quantum measurement}",
  journal = {Contemp. Phys.},
     year = 2006,
    month = sep,
   volume = 47,
    pages = {279-303},
      doi = {10.1080/00107510601101934},
   adsurl = {http://adsabs.harvard.edu/abs/2006ConPh..47..279J}
}

@article{PhysRevX.6.011002,
  title = {Observing Quantum State Diffusion by Heterodyne Detection of Fluorescence},
  author = {Campagne-Ibarcq, P. and Six, P. and Bretheau, L. and Sarlette, A. and Mirrahimi, M. and Rouchon, P. and Huard, B.},
  journal = {Phys. Rev. X},
  volume = {6},
  issue = {1},
  pages = {011002},
  numpages = {7},
  year = {2016},
  month = {Jan},
  publisher = {American Physical Society},
  doi = {10.1103/PhysRevX.6.011002},
  url = {https://link.aps.org/doi/10.1103/PhysRevX.6.011002}
}

@book{Bookwiseman2009,
  title={Quantum measurement and control},
  author={Wiseman, H. M. and Milburn, G. J.},
  year={2009},
  publisher={Cambridge University press},
  url = "https://www.cambridge.org/core/books/quantum-measurement-and-control/F78F445CD9AF00B10593405E9BAC6B9F"
}

@book{Bookjacobs2014,
  title={Quantum measurement theory and its applications},
  author={Jacobs, K.},
  year={2014},
  publisher={Cambridge University Press},
  url = "https://www.cambridge.org/core/books/quantum-measurement-theory-and-its-applications/120E32FFBEBF6EE0F6EC6F84D51DC907"
}

@book{Timereversal,
  title={The physics of time reversal},
  author={Sachs, Robert G},
  year={1987},
  publisher={University of Chicago Press}
}

@book{jacobs2010stochastic,
  title={Stochastic processes for physicists: understanding noisy systems},
  author={Jacobs, Kurt},
  year={2010},
  publisher={Cambridge University Press}
}

@article{Miguel1,
  title = {Resetting Uncontrolled Quantum Systems},
  author = {Navascu\'es, Miguel},
  journal = {Phys. Rev. X},
  volume = {8},
  issue = {3},
  pages = {031008},
  numpages = {10},
  year = {2018},
  month = {Jul},
  publisher = {American Physical Society},
  doi = {10.1103/PhysRevX.8.031008},
  url = {https://link.aps.org/doi/10.1103/PhysRevX.8.031008}
}

@article{Miguel2,
  title = {Universal Quantum Rewinding Protocol with an Arbitrarily High Probability of Success},
  author = {Trillo, D. and Dive, B. and Navascu\'es, M.},
  journal = {Phys. Rev. Lett.},
  volume = {130},
  issue = {11},
  pages = {110201},
  numpages = {5},
  year = {2023},
  month = {Mar},
  publisher = {American Physical Society},
  doi = {10.1103/PhysRevLett.130.110201},
  url = {https://link.aps.org/doi/10.1103/PhysRevLett.130.110201}
}

@article{Miguel3,
author = {P. Schiansky and T. Str\"{o}mberg and D. Trillo and V. Saggio and B. Dive and M. Navascu\'{e}s and P. Walther},
journal = {Optica},
number = {2},
pages = {200--205},
publisher = {Optica Publishing Group},
title = {Demonstration of universal time-reversal for qubit processes},
volume = {10},
month = {Feb},
year = {2023},
url = {https://opg.optica.org/optica/abstract.cfm?URI=optica-10-2-200},
doi = {10.1364/OPTICA.469109}
}

@inproceedings{arrighi,
  title={A toy model provably featuring an arrow of time without past hypothesis},
  author={Arrighi, Pablo and Dowek, Gilles and Durbec, Am{\'e}lia},
  booktitle={International Conference on Reversible Computation},
doi={10.1007/978-3-031-62076-8_4},
  pages={50--68},
  year={2024},
  organization={Springer}
}

@article{gravityarrow,
  title = {Identification of a Gravitational Arrow of Time},
  author = {Barbour, Julian and Koslowski, Tim and Mercati, Flavio},
  journal = {Phys. Rev. Lett.},
  volume = {113},
  issue = {18},
  pages = {181101},
  numpages = {5},
  year = {2014},
  month = {Oct},
  publisher = {American Physical Society},
  doi = {10.1103/PhysRevLett.113.181101},
  url = {https://link.aps.org/doi/10.1103/PhysRevLett.113.181101}
}

@article{PRLReversing,
  title = {Reversing Unknown Qubit-Unitary Operation, Deterministically and Exactly},
  author = {Yoshida, Satoshi and Soeda, Akihito and Murao, Mio},
  journal = {Phys. Rev. Lett.},
  volume = {131},
  issue = {12},
  pages = {120602},
  numpages = {6},
  year = {2023},
  month = {Sep},
  publisher = {American Physical Society},
  doi = {10.1103/PhysRevLett.131.120602},
  url = {https://link.aps.org/doi/10.1103/PhysRevLett.131.120602}
}

@article{Minev_2019,
   title={To catch and reverse a quantum jump mid-flight},
   volume={570},
   ISSN={1476-4687},
   url={http://dx.doi.org/10.1038/s41586-019-1287-z},
   DOI={10.1038/s41586-019-1287-z},
   number={7760},
   journal={Nature},
   publisher={Springer Science and Business Media LLC},
   author={Minev, Z. K. and Mundhada, S. O. and Shankar, S. and Reinhold, P. and Gutiérrez-Jáuregui, R. and Schoelkopf, R. J. and Mirrahimi, M. and Carmichael, H. J. and Devoret, M. H.},
   year={2019},
   month=jun, pages={200–204} }

@article{manzano2022quantum,
    author = {Manzano, Gonzalo and Zambrini, Roberta},
    title = "{Quantum thermodynamics under continuous monitoring: A general framework}",
    journal = {AQS},
    volume = {4},
    number = {2},
    pages = {025302},
    year = {2022},
    month = {05},
    issn = {2639-0213},
    doi = {10.1116/5.0079886}
}

@article{Mitchison2021chargingquantum,
  doi = {10.22331/q-2021-07-13-500},
  url = {https://doi.org/10.22331/q-2021-07-13-500},
  title = {Charging a quantum battery with linear feedback control},
  author = {Mitchison, Mark T. and Goold, John and Prior, Javier},
  journal = {{Quantum}},
  issn = {2521-327X},
  publisher = {{Verein zur F{\"{o}}rderung des Open Access Publizierens in den Quantenwissenschaften}},
  volume = {5},
  pages = {500},
  month = jul,
  year = {2021}
}

@article{PhysRevA.106.042221,
  title = {Thermodynamics of quantum measurement and {Maxwell's} demon's arrow of time},
  author = {Yanik, Kagan and Bhandari, Bibek and Manikandan, Sreenath K. and Jordan, Andrew N.},
  journal = {Phys. Rev. A},
  volume = {106},
  issue = {4},
  pages = {042221},
  numpages = {8},
  year = {2022},
  month = {Oct},
  publisher = {American Physical Society},
  doi = {10.1103/PhysRevA.106.042221},
  url = {https://link.aps.org/doi/10.1103/PhysRevA.106.042221}
}

@article{ArrowDresselPRL2017,
  title = {Arrow of Time for Continuous Quantum Measurement},
  author = {Dressel, Justin and Chantasri, Areeya and Jordan, Andrew N. and Korotkov, Alexander N.},
  journal = {Phys. Rev. Lett.},
  volume = {119},
  issue = {22},
  pages = {220507},
  numpages = {6},
  year = {2017},
  month = {Dec},
  publisher = {American Physical Society},
  doi = {10.1103/PhysRevLett.119.220507},
  url = {https://link.aps.org/doi/10.1103/PhysRevLett.119.220507}
}

@article{ArrowBigelowNatComm2021,
  title={Quantum measurement arrow of time and fluctuation relations for measuring spin of ultracold atoms},
  author={Jayaseelan, Maitreyi and K Manikandan, Sreenath and Jordan, Andrew N and Bigelow, Nicholas P},
  journal={Nat. Commun.},
  volume={12},
  doi = {10.5281/zenodo.4524924},
  number={1},
  pages={1--7},
  year={2021},
  publisher={Nature Publishing Group}
}

@article{ArrowMurchPRL2019,
  title = {Characterizing a Statistical Arrow of Time in Quantum Measurement Dynamics},
  author = {Harrington, P. M. and Tan, D. and Naghiloo, M. and Murch, K. W.},
  journal = {Phys. Rev. Lett.},
  volume = {123},
  issue = {2},
  pages = {020502},
  numpages = {6},
  year = {2019},
  month = {Jul},
  publisher = {American Physical Society},
  doi = {10.1103/PhysRevLett.123.020502},
  url = {https://link.aps.org/doi/10.1103/PhysRevLett.123.020502}
}

@article{MEnginePRL2017,
  title = {Extracting Work from Quantum Measurement in {Maxwell's} Demon Engines},
  author = {Elouard, Cyril and Herrera-Mart\'{\i}, David and Huard, Benjamin and Auff\`eves, Alexia},
  journal = {Phys. Rev. Lett.},
  volume = {118},
  issue = {26},
  pages = {260603},
  numpages = {6},
  year = {2017},
  month = {Jun},
  publisher = {American Physical Society},
  doi = {10.1103/PhysRevLett.118.260603},
  url = {https://link.aps.org/doi/10.1103/PhysRevLett.118.260603}
}

@article{MEnginePRL2018,
  title = {Efficient Quantum Measurement Engines},
  author = {Elouard, Cyril and Jordan, Andrew N.},
  journal = {Phys. Rev. Lett.},
  volume = {120},
  issue = {26},
  pages = {260601},
  numpages = {5},
  year = {2018},
  month = {Jun},
  publisher = {American Physical Society},
  doi = {10.1103/PhysRevLett.120.260601},
  url = {https://link.aps.org/doi/10.1103/PhysRevLett.120.260601}
}

@article{elouard2025revealing,
  title={Revealing the fuel of a quantum continuous measurement-based refrigerator},
  doi = {10.48550/arXiv.2502.10349},
  author={Elouard, Cyril and Manikandan, Sreenath K and Jordan, Andrew N and Haack, Geraldine},
  journal={arXiv preprint arXiv:2502.10349},
  year={2025}
}

@article{dassonneville,
      title={Amplifying microwave pulses with a single qubit engine fueled by quantum measurements}, 
      author={Dassonneville, R{\'e}my and Elouard, C and Cazali, Romain and Assouly, R and Bienfait, A and Auff{\`e}ves, A and Huard, B},
      doi = {10.48550/arXiv.2501.17069},
      year={2025},
      eprint={2501.17069},
      journal={arXiv}
}

@article{PhysRevE.105.044137,
  title = {Efficiently fueling a quantum engine with incompatible measurements},
  author = {Manikandan, Sreenath K. and Elouard, Cyril and Murch, Kater W. and Auff\`eves, Alexia and Jordan, Andrew N.},
  journal = {Phys. Rev. E},
  volume = {105},
  issue = {4},
  pages = {044137},
  numpages = {10},
  year = {2022},
  month = {Apr},
  publisher = {American Physical Society},
  doi = {10.1103/PhysRevE.105.044137},
  url = {https://link.aps.org/doi/10.1103/PhysRevE.105.044137}
}

\widetext
\clearpage
\appendix 
 
 \part*{\begin{center}
\normalsize{APPENDIX  
 } 
 \end{center}
 }

 \section{Dynamics of monitored quantum systems}
\label{app-stratonovich}

This Appendix derives a master equation for the dynamics of a monitored quantum system from the Kraus operators that describe generalized measurements. Throughout all appendices, we will use $\rho$ to denote a system's density matrix. We will assume pure states, $\rho^2 = \rho$. 

 The Kraus operator for the measurement of a Hermitian operator $A$ with outcome $r_t$ during a time-interval $\dt$ is
 \begin{align}
 \label{eq-app:Kraus}
 M_{r_t} = \mathcal{N} e^{- 2 \kappa \dt \left( r_t-A \right)^2} = \mathcal{N} e^{-2 \kappa \dt r_t^2 - 2 \kappa A^2 \dt  } e^{4 \kappa \dt r_t A},
 \end{align}
 where $\mathcal{N}\coloneqq \left( \frac{4 \kappa \dt }{ \pi} \right)^{1/4}$ is a normalization factor 
 and we defined $\kappa \coloneqq \frac{1}{8 \tau}$.
 In the limit of small $\dt$, 
 \begin{align}
 M_{r_t} \approx \mathcal{N} e^{-2 \kappa  r_t^2 \dt } \left( \id + 4\kappa r_t A \dt -2 \kappa A^2 \dt\right),
 \end{align}
 where we only keep track of the terms in the exponential that depend on the observable since all terms independent of $A$ cancel with the normalization $\mathcal{N}$. 
The measurement output tracks the observable's expectation value, 
\begin{align}
\label{eq-app:output}
    r_tdt = \langle A \rangle_t dt + \sqrt{\tau}dW_t,
\end{align}
where $dW_t$ is a stochastic zero-mean variable with $dW_t^2 = dt$~\cite{JacobsIntro2006}.

 Then, to first order, the change in the state becomes
 \begin{align}
 \label{eq-app:Strato1}
 \rho_{t+dt} &= \rho_t +  \frac{\left( \id + 4\kappa r_t A \dt -2 \kappa A^2 \dt\right) \rho_t \left( \id + 4\kappa r_t A \dt -2 \kappa A^2 \dt\right)}{\tr{   \left( \id + 4\kappa r_t A \dt  -2 \kappa A^2 \dt\right) \rho_t \left( \id + 4\kappa r A \dt -2 \kappa A^2 \dt\right) }}   \nonumber \\
 &\approx \rho_t  +  \frac{ \rho_t + 2 \kappa  \big\{ \rho_t, 2r_tA -A^2 \big\} dt  }{\tr{  \rho_t + 2 \kappa  \big\{ \rho, 2r_tA -A^2\big\} dt }}       \nonumber \\
 &\approx \rho_t + \Big( \rho_t + 2 \kappa  \big\{ \rho_t, 2r_tA -A^2  \big\} dt \Big) \Big( 1 - 2\kappa   4 r_t\langle A \rangle_t  \dt + 2\kappa 2 \langle A^2 \rangle_t \dt \Big) \nonumber \\
 &\approx  \rho_t + 4\kappa r_t \Big( \{ \rho_t,A \}  - 2 \langle A \rangle_t \rho_t \Big)    \dt - 2\kappa \Big( \{ \rho_t,A^2 \}  - 2 \langle A^2 \rangle_t \rho_t \Big)    \dt\nonumber \\
 &= \rho_t + \frac{r_t}{2\tau} \Big( \{ \rho_t,A \}  - 2 \langle A \rangle_t \rho_t \Big)    \dt - \frac{1}{4\tau} \Big( \{ \rho_t,A^2 \}  - 2 \langle A^2 \rangle_t \rho_t \Big) \dt.
 \end{align}
 This is valid for continuous Gaussian measurements. When the system is also subjected to a Hamiltonian $H$, a similar calculation shows that
\begin{align}
 \label{eq-app:Strato2}
 d\rho_t &= -i[H,\rho_t]dt + \frac{r_t}{2\tau} \Big( \{ \rho_t,A \}  - 2 \langle A \rangle_t \rho_t \Big)    \dt - \frac{1}{4\tau} \Big( \{ \rho_t,A^2 \}  - 2 \langle A^2 \rangle_t \rho_t \Big) \dt.
 \end{align}
This proves Eq.~\eqref{eq:Strato} in the main text. 

 Equation~\eqref{eq-app:Strato2} should be interpreted in the Stratonovich picture of stochastic calculus~\cite{GisinStrato}. It determines the state update with a midpoint increment evaluation, 
 \begin{align}
 \label{eq-app:StratoUpdate}
   d\rho_t \equiv 2 \big( \rho_{t+dt/2} -\rho_t \big),  
 \end{align}
obtained from adopting the convention $\rho_{t+dt/2} \nobreak=\nobreak \big( \rho_t \nobreak+\nobreak \rho_{t+dt} \big)/2 \nobreak=\nobreak \rho_t \nobreak+\nobreak d\rho_t/2$ when integrating~\cite{gardinerstochastic,jacobs2010stochastic,jackson2023simultaneous}.
Stratonovich integrals are $\int_0^T  f_t r_t dt \equiv \lim_{dt \rightarrow 0} \sum_{j=1}^N   f_{t_{j} + \tfrac{dt}{2}}   r_{t_j} dt \equiv  \lim_{dt \rightarrow 0} \sum_{j=1}^N \frac{   f_{t_{j+1}} + f_{t_j} }{2} r_{t_j} dt$. 
As a result of the update rule used in the Stratonovich picture, functions $f(t)$ are not independent of the white noise $dW_t$ in the measurement output, $r_tdt = \langle A \rangle_t + \sqrt{\tau}dW_t$~\cite{gardinerstochastic}.

For completeness, note that, in the It\^o picture, Eq.~\eqref{eq-app:Strato2} becomes~\cite{JacobsIntro2006}
\begin{align}
\label{eq-app:Ito}
d_{\textnormal{It\^o}}\rho_t = -i[H,\rho_t]dt - \frac{1}{8\tau} [A,[A,\rho_t]] dt + \sqrt{\frac{1}{4\tau}} \big( \{A,\rho_t \} - 2\langle A \rangle_t \rho_t \big) dW_t.
\end{align}
See Ref.~\cite{Bookwiseman2009} for a detailed guide on transforming between the It\^o and Stratonovich pictures, and the advantages and disadvantages of each one. In particular, taking ensemble averages is more straightforward in It\^o, and averaging Eq.~\eqref{eq-app:Ito} recovers a Lindblad master equation for the ensemble-averaged state $\overline{\rho_t}$.  
In contrast, the Stratonovich picture more accurately reflects the dynamics of a system subject to physical stochastic noise which, under idealized limits, becomes a Wiener process~\cite{Bookwiseman2009}.

\section{Replicating stochastic quantum trajectories}
\label{app-replicating}
In this Appendix, we prove that the Hamiltonian $\Hmeas$, given by Eq.~\eqref{eq:Hmeas} in the main text, reproduces the stochastic dynamics of a monitored quantum system.

Driven by the Hamiltonian
\begin{align}
\label{eq-app:Hmeas}
\Hmeas \coloneqq -i \frac{r_t}{2\tau} \left[\rho_t,A \right] +i \frac{1}{4\tau} \left[\rho_t,A^2 \right], 
\end{align} 
the state evolves according to
\begin{align}
\label{eq-app:drhofeedback}
d\rho_t &= -i \left[ \Hmeas,\rho_t\right]\dt = - \frac{r_t}{2\tau} \left[ \left[ \rho_t,  A \right], \rho_t \right] \dt + \frac{1}{4\tau} \left[ \left[ \rho_t,  A^2 \right], \rho_t \right] \dt \nonumber \\
&= - \frac{r_t}{2\tau}  \bigg( \rho_t  A  \rho_t -   A  \rho_t - \rho_t   A   + \rho_t   A   \rho_t    \bigg) \dt + \frac{1}{4\tau}  \bigg( \rho_t  A^2  \rho_t -   A^2  \rho_t - \rho_t   A^2   + \rho_t   A^2   \rho_t    \bigg) \dt\nonumber \\
&= \frac{r_t}{2\tau} \bigg( \{ \rho_t,A \} - 2\langle A \rangle \rho_t \bigg) \dt - \frac{1}{4\tau} \bigg( \{ \rho_t,A^2 \} - 2\langle A^2 \rangle \rho_t \bigg) \dt,
\end{align} 
where we used the fact that $\rho_t^2 = \rho_t$ is pure. Equation~\eqref{eq-app:drhofeedback} is identical to Eq.~\eqref{eq-app:Strato1}. $\Hmeas$ thus retraces the dynamics of a monitored quantum system, as claimed in the main text. This is confirmed by numerical simulations, which show that the dynamics under Eq.~\eqref{eq-app:Hmeas} equals the dynamics under a sequence of Kraus operators~\eqref{eq-app:Kraus}.

\section{The length of the quantum arrow of time}
\label{app-quantumarrow}

Here, we show that a quantifiable arrow of time emerges in monitored quantum systems, as first considered in Ref.~\cite{ArrowDresselPRL2017}. 
We reproduce the main results of Ref.~\cite{ArrowDresselPRL2017}, generalizing them from a single qubit to arbitrary systems for any $A$ such that $A^2 = \id$.
We prove Eq.~\eqref{eq:arrow} in the main text.

The probability of obtaining an outcome $r_t$ when performing a generalized measurement on a state $\rho_t$ is
\begin{align}
 \label{eq-app:probsingle}
 P(r_t|\rho_t) \nobreak  = \tr{\rho_t M_{r_t}^\dag M_{r_t}} = \nobreak \sqrt{\frac{dt}{2\pi\tau}} \exp \left[ -\frac{\left(r_t-\langle A \rangle \right)^2 dt}{ 2\tau } \right].
 \end{align}

The forward and backward processes of a monitored system with a Hamiltonian $H$ are
 \begin{align} 
 \label{eq-app:forward}
 \ket{\psi}_{T} = \frac{ M_{r_N} e^{-i H dt} \ldots 
\overbrace{M_{r_j}}^{\color{blue}\text{\clap{outcomes track state $\ket{\psi}_{t_j}$ that comes from our past}}}
 e^{-i H dt} \ldots M_{r_0} e^{-i H dt} \ket{\psi}_{0} }{\sqrt{\Prob_F(\rvec)}}
 \end{align}
and
 \begin{align}
 \label{eq-app:backward}
 \ket{\psi}_{0} = \frac{ M_{-r_0} e^{-i \widetilde{H} dt} \ldots 
 e^{-i \widetilde{H} dt} \overbrace{M_{-r_j}}^{\color{red}\text{\clap{outcomes track state $\ket{\psi}_{t_{j+1}}$ that comes from our \emph{future}}}}
  \ldots e^{-i \widetilde{H} dt}  M_{-r_N}  \ket{\psi}_{T}   }{\sqrt{\Prob_B(\rvec)}}.
 \end{align}

The probabilities of an outcome $r_j$ in the forward process and the corresponding outcome $-r_j$ in the backward process can be obtained from Eq.~\eqref{eq-app:probsingle}. Crucially, note that $r_j$ in the forward process follows a state $\ket{\psi}_{t_{j}}$, but $-r_{j}$ in the backward process follows the state $\ket{\psi}_{t_{j+1}}$. 
This leads to the following expressions for the probability of the forward and backward processes:
\begin{subequations}
\label{eq-app:probfwdback}
\begin{align}
\label{eq-app:forwardprob}
\Prob_F(\rvec) &\equiv \prod_{j=0}^N P\big(r_{t_j}\big\vert\ket{\psi}_{t_{j}}\!\big) =   \prod_{j=0}^N \sqrt{\frac{dt}{2\pi\tau}} \exp \left[ -\big( r_{t_j}-\langle A \rangle_{t_{j}} \big)^2 \frac{dt}{ 2\tau } \right], \\
\label{eq-app:backwardprob}
\Prob_B(\rvec) &\equiv \prod_{j=0}^N P\big(-r_{t_j}\big\vert\ket{\psi}_{t_{j+1}}\!\big) =   \prod_{j=0}^N \sqrt{\frac{dt}{2\pi\tau}} \exp \left[ -\big( -r_{t_j}-\langle A \rangle_{t_{j+1}} \big)^2 \frac{dt}{ 2\tau } \right],
\end{align}
\end{subequations}
where we used Eq.~\eqref{eq-app:probsingle}. 

Then, 
\begin{align}
\label{eq-app:arrow}
\ln \mathcal{R}  &\coloneqq \ln \frac{\Prob_F(\rvec)}{\Prob_B(\rvec)} = \sum_{j=0}^N 2 r_{t_j} \big( \langle A \rangle_{t_{j+1}} + \langle A \rangle_{t_j}  \big) \frac{dt}{2\tau}     \equiv \frac{2}{\tau} \int_0^T r_t \langle A \rangle_t  \dt.
\end{align}
The final integral should be interpreted in the Stratonovich picture, with midpoint integration~\cite{jacobs2010stochastic}. This proves Eq.~\eqref{eq:arrow} in the main text.

From Eq.~\eqref{eq-app:Strato2}, the system's state after observing a measurement outcome $r_{t_j}$ is
\begin{align}
    \rho_{t_{j+1}} = \rho_{t_{j}} + d\rho = \rho_{t_{j}} -i[H,\rho_{t_{j}}]dt + \frac{r_{t_j}}{2\tau} \Big( \{ \rho_{t_{j}},A \}  - 2 \langle A \rangle_{t_{j}} \rho_{t_{j}} \Big)    \dt.
\end{align}
Then, the observable's expectation value becomes
\begin{align}
    \langle A \rangle_{t_{j+1}} &= \langle A \rangle_{t_{j}} - i \langle [A,H ] \rangle_{t_{j}} dt + \frac{r_{t_j} dt}{2\tau} \tr{A \Big( \{ \rho_{t_{j}},A \}  - 2 \langle A \rangle_{t_{j}} \rho_{t_{j}} \Big)} \nonumber \\
    &= \langle A \rangle_{t_{j}} - i \langle [A,H ] \rangle_{t_{j}} dt + \frac{r_{t_j} dt}{2\tau} \left( 2 - 2\langle A \rangle^2_{t_{j}}\right).
\end{align}
The summands in Eq.~\eqref{eq-app:arrow} thus are
\begin{align}
\label{eq-app:arrowV2}
    2 r_{t_j} \big( \langle A \rangle_{t_{j+1}} + \langle A \rangle_{t_j}  \big) \frac{dt}{2\tau} &= \frac{r_{t_j} dt}{\tau} \left(    2\langle A \rangle_{t_{j}} - i \langle [A,H ] \rangle_{t_{j}} dt + \frac{r_{t_j} dt}{\tau} \left( 1 - \langle A \rangle^2_{t_{j}}\right)   \right) \nonumber \\
    &\approx \frac{r_{t_j} dt}{\tau} 2\langle A \rangle_{t_{j}} + 
    \left(\frac{r_{t_j} dt}{\tau} \right)^2 \left( 1 - \langle A \rangle^2_{t_{j}}\right)    \nonumber \\
    &= 2\frac{r_{t_j} dt}{\tau} \langle A \rangle_{t_{j}} + 
     \frac{dt}{\tau} \left( 1 - \langle A \rangle^2_{t_{j}}\right).
\end{align}
In the 3rd line, we used the fact that $(r_jdt)^2 = \tau dt$ to leading order. This holds from the fact that  $r_t dt \nobreak =\nobreak \langle A \rangle_t dt \nobreak+\nobreak  \sqrt{\tau} dW_t$ and that $dW_t^2 = dt$ in stochastic calculus~\citep{jacobs2010stochastic}. 

Inserting Eq.~\eqref{eq-app:arrowV2} into Eq.~\eqref{eq-app:arrow}, and upon averaging over realizations, we obtain
\begin{align}
\label{eq-app:averagedArrow}
\overline{ \ln \mathcal{R} }  &= \sum_{j=0}^N 
  \overline{ 2\frac{r_{t_j} dt}{\tau} \langle A \rangle_{t_{j}} + 
     \frac{dt}{\tau} \left( 1 - \langle A \rangle^2_{t_{j}}\right)} = \sum_{j=0}^N 
   2\frac{ \overline{\langle A \rangle_{t_{j}}^2} dt}{\tau}  + 
     \frac{dt}{\tau} \left( 1 - \overline{\langle A \rangle_{t_{j}}^2}\right)  \nonumber \\
     &= \sum_{j=0}^N 
     \frac{dt}{\tau} \left( 1 + \overline{\langle A \rangle_{t_{j}}^2}\right) = \int_0^T \frac{1+ \overline{\langle A \rangle_t^2}}{\tau} dt. 
\end{align}
In the first line, we used that $\overline{ r_{t_j} \langle A \rangle_{t_{j}}} = \overline{ \langle A \rangle_{t_{j}}^2}$, which holds since $dW_t$ is a Wiener process~\cite{jacobs2010stochastic}. This proves a claim made before Eq.~\eqref{eq:arrow-steeredSimple} in the main text and proves the main results of Ref.~\cite{ArrowDresselPRL2017} for any $A$ such that $A^2 = \id$. For a qubit with $H=\omega/2\sigma_y$ and $A=\sigma_z$, $\langle A \rangle_t$ will, on average, perform Rabi oscillations~\cite{ArrowDresselPRL2017}, so $\int_0^T \overline{\langle A \rangle_t^2} dt/T \approx 1/2$, which yields $\overline{\ln \mathcal R} \approx 3T/(2\tau)$.

\section{Feedback dynamics}
\label{app-feedback}
Here, we derive the dynamics of a system driven by a feedback Hamiltonian proportional to $\Hmeas$. We leverage such feedback dynamics to modify the quantum arrow of time in the next appendix. We consider $A^2 = \id$ from now on, as reversed trajectories are possible physical processes in such a case. 

The Hamiltonian $\Hmeas$ in Eq.~\eqref{eq-app:Hmeas} reproduces the dynamics of a monitored quantum system. Here, we consider the dynamics of a monitored system that exploits a Hamiltonian of the form~\eqref{eq-app:Hmeas} to perform feedback.
We consider a feedback Hamiltonian 
\begin{align}
\label{eq-app:Hfeedback}
\Hfb \coloneqq \chron \Hmeas
= -i \chron \frac{r_t}{2\tau} \left[\rho_t,A \right].
\end{align} 

Note that the feedback Hamiltonian acts after the system's state has changed due to a measurement instance. In other words, feedback with a measurement outcome $r_t$ acts \emph{after} the state has been affected by the observation of $r_t$. We can make this explicit  by considering discrete infinitesimal changes and writing Eq.~\eqref{eq-app:Hfeedback} as
\begin{align}
\label{eq-app:HfeedbackDiscrete}
\Hfb \coloneqq \chron \Hmeas 
= -i \chron \frac{r_{t_{j-1}}}{2\tau} \left[\rho_{t_j},A \right].
\end{align}

Thus, the change of a state $\rho_t$ can be obtained by combining Eq.~\eqref{eq-app:Strato2}, which accounts for the measurement back-action plus Hamiltonian dynamics, with the dynamics generated by Eq.~\eqref{eq-app:HfeedbackDiscrete} from the previous feedback instance (since that is the one that affects the state at time $t$). Thus, we have
\begin{align}
    \label{eq-app:HfeedbackV2}
   d\rho_{t_j} &= -i[H,\rho_{t_j}]dt + \frac{r_{t_{j}}}{2\tau} \Big( \{ \rho_{t_j},A \}  - 2 \langle A \rangle_{t_j} \rho_{t_j} \Big)    \dt +   \chron \frac{r_{t_{j-1}}}{2\tau} \Big( \{ \rho_{t_j},A \}  - 2 \langle A \rangle_{t_j} \rho_{t_j} \Big) \dt.
\end{align}
Equation~\eqref{eq-app:HfeedbackV2} assumes the minimum possible physical feedback delay, where $\Hfb$ acts immediately after a measurement instance. In practice, one would have a feedback delay timescale $\tau > dt$, so the feedback $\Hfb$ would depend on $r_{t - \tau}$. Equation~\eqref{eq-app:HfeedbackV2} is a good approximation if the feedback delay is smaller than all other timescales relevant to the problem (e.g., frequencies in $H$ and $\tau$).
 
Using Eq.~\eqref{eq-app:drhofeedback}, we can write~\eqref{eq-app:HfeedbackV2} as
\begin{align}
    \label{eq-app:HfeedbackV2Aux}
   d\rho_{t_j} &= -i[H,\rho_{t_j}]dt -i[\Hmeas,\rho_{t_j}]dt -i[\Hfb,\rho_{t_j}]dt.
\end{align}

\section{Reshaping the quantum arrow of time}
\label{app-steeringarrow}

Next, we prove that feedback dynamics with a Hamiltonian $\chron\Hmeas$ can modify the quantum arrow of time. The parameter $\chron$ will characterize the direction and magnitude by which the arrow of time changes. We prove Eqs.~\eqref{eq:steeredarrow} and~\eqref{eq:arrow-steeredSimple} in the main text.

To study $\ln \mathcal{R}_\chron$, we revisit Eq.~\eqref{eq-app:arrow} with the feedback dynamics:
\begin{align}
\label{eq-app:arrowV2chron}
\ln \mathcal{R}_\chron  &\coloneqq \ln \frac{\Prob_F^\chron(\rvec)}{\Prob_B^\chron(\rvec)} = \sum_{j=0}^N 2 r_{t_j} \big( \langle A \rangle_{t_{j+1}}' + \langle A \rangle_{t_j} \big) \frac{dt}{2\tau}.
\end{align}

Unlike the expression in~\eqref{eq-app:arrow}, $\langle A \rangle_{t_{j+1}}'$ in Eq.~\eqref{eq-app:arrowV2chron} involves a feedback step with $\chron\Hmeas$, where we use $\rho_{t_j}'$ to denote the state affected by the feedback with $\chron\Hmeas$. The state's change due to feedback is described by the last term in Eq.~\eqref{eq-app:HfeedbackV2}, so
\begin{align}
\label{eq-app:changeChron}
\rho_{t_{j+1}}' &= \rho_{t_{j+1}} + \frac{\chron r_{t_j} dt}{2\tau} \bigg( \{ \rho_{t_{j+1}},A \} - 2\langle A \rangle_{t_{j+1}} \rho_{t_{j+1}} \bigg).
\end{align} 
The feedback's effect on the observable's expectation value thus is
\begin{align}
\langle A \rangle_{t_{j+1}}' &= 
\tr{A\rho_{t_{j+1}}'} = \langle A \rangle_{t_{j+1}} + \chron \frac{ r_{t_{j}}  }{2\tau} \bigg( \tr{ A\{ \rho_{t_{j+1}},A \}} - 2\langle A \rangle_{t_{j+1}} \tr{A\rho_{t_{j+1}}} \bigg) \dt \nonumber \\
&= \langle A \rangle_{t_{j+1}} + \chron \frac{ r_{t_{j}}  }{2\tau} \bigg( 2 \tr{ \rho_{t_{j+1}} A^2 } - 2\langle A \rangle_{t_{j+1}}^2\bigg) \dt \nonumber \\
&= \langle A \rangle_{t_{j+1}} + \chron \frac{ r_{t_{j}}  }{\tau} \Big( 1 - \langle A \rangle_{t_{j+1}}^2\Big) \dt,
\end{align} 
where we used that $A^2 = \id$.
 
Inserting $\langle A  \rangle_{t_{j+1}}'$ into Eq.~\eqref{eq-app:arrowV2chron} gives
\begin{align}
\ln \mathcal{R}_\chron &= \sum_{j=0}^N 2 r_{t_j} \big( \langle A \rangle_{t_{j+1}}' + \langle A \rangle_{t_{j}}  \big) \frac{dt}{2\tau} \nonumber \\
&=   \sum_{j=0}^N 2 r_{t_j} \left(  \langle A  \rangle_{t_{j+1}} + \langle A  \rangle_{t_{j}} + \chron \frac{ r_{t_{j}}  }{\tau} \Big( 1 - \langle A \rangle_{t_{j+1}}^2\Big) \dt \right) \frac{dt}{2\tau} \nonumber \\
&=   \ln \mathcal{R} + \chron \sum_{j=0}^N \Big( 1 - \langle A  \rangle_{t_{j+1}}^2\Big) \frac{  r_{t_j}^2 dt^2}{\tau^2} \nonumber \\
&= \ln \mathcal{R} + \chron \sum_{j=0}^N \Big( 1 - \langle A  \rangle_{t_{j+1}}^2 \Big) \frac{\tau dt}{\tau^2}
\nonumber \\
&= \ln \mathcal{R} +  \frac{\chron}{\tau}\int_0^T \! \big(1-\langle A \rangle_t^2 \big) \, dt,
\end{align}
which proves Eq.~\eqref{eq:steeredarrow} in the main text. $\ln \mathcal{R}$ in the third line is the feedback-less contribution to $\ln \mathcal{R}_\chron$, as in Eq.~\eqref{eq-app:arrow}.
In the 4th line, we used the fact that $(r -\langle A \rangle)^2 dt^2 = \tau dt$ to leading order. This holds because  $r_t dt \nobreak =\nobreak \langle A \rangle_t dt \nobreak+\nobreak  \sqrt{\tau} dW_t$ and because $dW_t^2 = dt$ in stochastic calculus~\citep{jacobs2010stochastic}.

Using Eq.~\eqref{eq-app:averagedArrow}, 
\begin{align}
\overline{ \ln \mathcal{R}_\chron } &= \overline{ \ln \mathcal{R} } + \frac{\chron }{\tau}\int_0^T \left( 1-\overline{\langle A \rangle_t^2} \right) dt = \overline{ \ln \mathcal{R} } + \chron \frac{T}{\tau} + \chron \frac{T}{\tau} - \chron \overline{\ln \mathcal{R}} \nonumber \\
&= (1-\chron) \, \overline{\ln \mathcal{R}} + 2 \frac{T}{\tau} \chron.
\end{align}
$\chron = 0$ recovers the standard arrow of time $\overline{\ln \mathcal{R}}$.

For a qubit with $H=\omega/2\sigma_y$ and $A=\sigma_z$, $\overline{\ln \mathcal R} \approx 3T/(2\tau)$ for long enough runtimes~\cite{ArrowDresselPRL2017}.
 Then, 
\begin{align}
\overline{\ln \mathcal{R}_\chron} \approx (1-\chron) \frac{3T}{2\tau} + 2 \frac{T}{\tau} \chron = \frac{T}{2\tau}\left( 3 - 3\chron +4\chron \right) = \frac{T}{2\tau}\left( 3 + \chron  \right).
\end{align}
This proves Eq.~\eqref{eq:arrow-steeredSimple} in the main text.

\section{A continuous measurement engine}
\label{sec-app:engine}
Here, we study the output of a continuous measurement engine that draws energy from the monitoring process by exploiting $\Hfb$. We prove Eqs.~\eqref{eq:work} and~\eqref{eq:workrate} in the main text. We also study the minimum thermodynamic cost to reset the memory of the agent performing feedback.

The rate at which a time-dependent Hamiltonian $\mathcal{H}_t$ does thermodynamic work on a system is
$\dot W \coloneqq \tr{\rho_t   \mathcal{\dot H}_t}$~\cite{Strasberg_2021,manzano2022quantum}. 
Thus, the work performed by the feedback Hamiltonian $\Hfb$ over an interval $dt$ after observing an outcome $r_{t_{j-1}}$ is
\begin{align}
\label{eq-app:work}
    \delta W_{t_j} &\coloneqq \tr{\rho_{t_j} d \Hfb} = -i \chron \frac{1}{2\tau} \tr{\rho_{t_j} d \Big( r_{t_{j-1}}[\rho_{t_j} , A] \Big)}  = -i \chron \frac{r_{t_{j-1}}}{2\tau} \tr{\rho_{t_j}  [d\rho_{t_j} , A] } \nonumber \\
    &= -i \chron \frac{r_{t_{j-1}}}{2\tau} \tr{  d\rho_{t_j} [A,\rho_{t_j}] } = -i \chron \frac{r_{t_{j-1}}dt}{2\tau} \tr{  -i[H,\rho_{t_j}] [A,\rho_{t_j}] } -i \chron \frac{r_{t_{j-1}} dt}{2\tau} \tr{  -i[\Hmeas + \Hfb, \rho_{t_j}] [A,\rho_{t_j}] } \nonumber \\
    &= \chron \frac{r_{t_{j-1}} dt}{2\tau} \tr{ [H,\rho_{t_j}] [\rho_{t_j},A] } \nonumber \\
    &= \chron \frac{r_{t_{j-1}} dt}{\tau} \cov_{t_j} (A,H).
\end{align}
The measurement output and state are at instances $t_{j-1}$ and $t_j$, respectively, because the feedback affects the state after a measurement instance has already affected it (see Sec.~\ref{app-feedback}). We used Eq.~\eqref{eq-app:HfeedbackDiscrete} and the cyclicity of the trace in the first line. We used the fact that $\tr{  -i[\Hmeas + \Hfb, \rho_{t_j}] [A,\rho_{t_j}] } = 0$ when going from the second to the third line, which follows from the definitions of $\Hmeas$ and $\Hfb$. The final expression follows from the definition of the covariance, $\cov_{t_j}(A,H) \nobreak\coloneqq\nobreak \langle \{A,H\} \rangle_{t_j}/2 \nobreak-\nobreak \langle A \rangle_{t_j} \langle H \rangle_{t_j}$. This proves Eq.~\eqref{eq:work} in the main text.

To evaluate the feedback's effect on the work rate, we use 
\begin{align}
  \label{eq-app:HfeedbackV2aux}
   \rho_{t_{j}} &= \rho_{t_{j-1}}-i[H,\rho_{t_{j-1}}]dt + \frac{r_{t_{j-1}}}{2\tau} \Big( \{ \rho_{t_{j-1}},A \}  - 2 \langle A \rangle_{t_{j-1}} \rho_{t_{j-1}} \Big)    \dt +   \chron \frac{r_{t_{j-2}}}{2\tau} \Big( \{ \rho_{t_{j-1}},A \}  - 2 \langle A \rangle_{t_{j-1}} \rho_{t_{j-1}} \Big) \dt,
\end{align}
from Eq.~\eqref{eq-app:HfeedbackV2}. This implies that, to leading order, the average work rate is 
\begin{align}
 \overline{\delta W_{t_j} }& = \chron \overline{ \frac{r_{t_{j-1}} dt}{\tau} \cov_{t_j}(A,H)}  \approx \chron \overline{ \frac{r_{t_{j-1}} dt}{\tau} \cov_{t_{j-1}}(A,H)} \nonumber \\
 &\qquad \qquad \qquad +  \frac{1}{2} 
 \chron \overline{ \left( \frac{r_{t_{j-1}} dt}{\tau} \right)^2 \frac{1}{2} \tr{  \{A,H\} \bigg( \{ \rho_{t-dt},A \} - 2\langle A \rangle_{t-dt } \rho_{t-dt} \bigg) } } \nonumber \\
  &\qquad\qquad\qquad- \frac{1}{2} \chron \overline{ \left( \frac{r_{t_{j-1}} dt}{\tau} \right)^2   
 \tr{  A \bigg( \{ \rho_{t_{j-1}},A \} - 2\langle A \rangle_{t_{j-1}} \rho_{t_{j-1}} \bigg) } \langle H \rangle_{t_{j-1}} } \nonumber \\
  &\qquad\qquad\qquad - \frac{1}{2} \chron \overline{ \left( \frac{r_{t_{j-1}} dt}{\tau} \right)^2     
  \tr{  H \bigg( \{ \rho_{t_{j-1}},A \} - 2\langle A \rangle_{t_{j-1}} \rho_{t_{j-1}} \bigg) } \langle A \rangle_{t_{j-1}}  } \nonumber \\
  &= \chron \overline{ \frac{r_{t_{j-1}} dt}{\tau} \cov_{t_{j-1}}(A,H)} + \frac{1}{2}
  \chron  \frac{dt}{\tau} \overline{ \Big( \langle H \rangle_{t_{j-1}}   +\langle A H A \rangle_{t_{j-1}}  - \langle A \rangle_{t_{j-1}} \langle \{A,H\} \rangle_{t_{j-1}}   \Big) }     \nonumber \\
  &\qquad\qquad - \frac{1}{2}  \chron  \frac{dt}{\tau} 2 \overline{ \left( 1 -\langle A \rangle_{t_{j-1}}^2 \right) \langle H \rangle_{t_{j-1}} }  -  \frac{1}{2} \chron  \frac{dt}{\tau}  2 \, \overline{ \cov_{t_{j-1}}(H,A) \langle A \rangle_{t_{j-1}} }.
\end{align}
When inserting Eq.~\eqref{eq-app:HfeedbackV2aux} into Eq.~\eqref{eq-app:work}, all other terms are of higher order in $dt$ (note that $\overline{r_{t_{j-2}} r_{t_{j-1}}} = \mathcal{O}(dt^2)$ since the white noise terms $dW_{t_{j-2}}$ and $dW_{t_{j-1}}$ are uncorrelated). We also used that $\overline{(r_t dt/\tau)^2} = dt/\tau$.

\subsection*{The engine's output for a qubit}
\label{sec-app:qubitEngine}

For a qubit with $H = \omega \sigma_y/2$ and $A = \sigma_z$, $\{A,H\}=0$ and $AHA = -H$, which gives
\begin{align}
\label{eq-app:workrateaverage}
 \overline{\delta W_{t_j} }& =   - \chron \overline{ \frac{r_{t_{j-1}} dt}{\tau} \langle A \rangle_{t_{j-1}} 
 \langle H \rangle_{t_{j-1}}  }     
 - \frac{1}{2} \chron  \frac{dt}{\tau} 2 \overline{ \left( 1 -\langle A \rangle_{t_{j-1}}^2 \right) \langle H \rangle_{t_{j-1}} }  + \frac{1}{2} \chron  \frac{dt}{\tau}  2 \, \overline{ \langle H \rangle_{t_{j-1}} \langle A \rangle^2_{t_{j-1}} }    \nonumber \\
 & =   - \chron \overline{ \frac{r_{t_{j-1}} dt}{\tau} \langle A \rangle_{t_{j-1}} 
 \langle H \rangle_{t_{j-1}}  }     
 - \frac{1}{2} \chron  \frac{dt}{\tau} 2 \overline{ \left( 1 - 2\langle A \rangle_{t_{j-1}}^2 \right) \langle H \rangle_{t_{j-1}} }.
\end{align}

For $\chron = -1$, the feedback counteracts the effect of the monitoring. Then, $\langle H \rangle$ remains constant, and $\langle A \rangle$ performs Rabi oscillations, with $\int_0^T \langle A \rangle^2 dt/T \approx 1/2$ when $T$ is large or a multiple of $2\pi/\omega$.
Moreover, $\overline{r_{t_{j-1}} \langle A \rangle_{t_{j-1}}} dt = \overline{  \langle A \rangle_{t_{j-1}}^2  } dt$ from Eq.~\eqref{eq-app:output}. 
Then, for $\chron = -1$, integrating Eq.~\eqref{eq-app:workrateaverage} over a time $T$ gives
\begin{align}
\label{eq-app:WorkOutput}
    \frac{\overline{W^\textnormal{out}_T}}{T} &\coloneqq - \frac{1}{T}\int_0^T \overline{\delta W_t} =  - \frac{1}{T}\sum_{j= 1}^N \overline{\delta W_{t_j}} 
 = - \frac{\langle H \rangle}{\tau} \int_0^T \langle A \rangle_t^2 \frac{dt}{T} -  \frac{\langle H \rangle}{\tau}   \int_0^T \overline{ \left( 1 - 2\langle A \rangle_t^2 \right)  } \frac{dt}{T}  
    = -\frac{\langle H \rangle}{2 \tau},
\end{align}
which proves Eq.~\eqref{eq:workrate} in the main text. Note the global minus sign in Eq.~\eqref{eq-app:WorkOutput}, which comes from focusing on the work extracted by the engine instead of the work done on the system [Eq.~\eqref{eq-app:workrateaverage}].

\end{document}